\documentclass[journal,10pt]{IEEEtran}
\usepackage{cite}
\usepackage{graphicx}
\usepackage{amsmath}
\usepackage{array}
\usepackage{mdwmath}
\usepackage{amssymb}
\usepackage{mdwtab}
\usepackage{stfloats}
\usepackage[tight,footnotesize]{subfigure}
\usepackage{amsmath,amsthm}
\usepackage{threeparttable}
\usepackage{color}
\usepackage{url}
\usepackage{algpseudocode}
\usepackage{algorithm}
\usepackage{multicol}
\usepackage{amssymb}
\usepackage{epstopdf}

\algrenewcommand{\algorithmicrequire}{\textbf{Input:}}
\algrenewcommand{\algorithmicensure}{\textbf{Output:}}

\newtheorem{theorem}{Theorem}
\newtheorem{remark}{Remark}

\begin{document}

\title{Robust Analysis of Full-Duplex Two-Way Space Shift Keying With RIS Systems}
\author{
Xusheng Zhu, Wen Chen, \IEEEmembership{Senior Member, IEEE}, Qingqing Wu, \IEEEmembership{Senior Member, IEEE}, Wen Fang, \\ Chaoying Huang, and Jun Li, \IEEEmembership{Senior Member, IEEE}

\thanks{
(\emph{Corresponding author: Wen Chen.)}}
\thanks{X. Zhu, W. Chen, Q. Wu, W. Fang, and C. Huang are with the Department of Electronic Engineering, Shanghai Jiao Tong University, Shanghai 200240, China (e-mail: xushengzhu@sjtu.edu.cn; wenchen@sjtu.edu.cn; qingqingwu@sjtu.edu.cn; wendyfang@sjtu.edu.dn; chaoyinghuang@sjtu.edu.cn).}
\thanks{J. Li is with the School of Electronic and Optical Engineering,
Nanjing University of Science Technology, Nanjing 210094, China (e-mail:
jun.li@njust.edu.cn).}
}

\maketitle
\begin{abstract}
Reconfigurable intelligent surface (RIS)-assisted index modulation system schemes are considered to be a promising technology for sixth-generation (6G) wireless communication systems, which can enhance various system capabilities such as coverage and reliability. However, obtaining perfect channel state information (CSI) is challenging due to the lack of a radio frequency chain in RIS.
{In this paper, we investigate the RIS-assisted full-duplex (FD) two-way space shift keying (SSK) system under imperfect CSI, where the signal emissions are augmented by deploying RISs in the vicinity of two FD users.
The maximum likelihood detector is utilized to recover the transmit antenna index.
With this in mind, we derive closed-form average bit error probability (ABEP) expression based on the Gaussian-Chebyshev quadrature (GCQ) method, and provide the upper bound and asymptotic ABEP expressions in the presence of channel estimation errors.}
To gain more insights, we also derive the outage probability and provide the throughput of the proposed scheme with imperfect CSI.
The correctness of the analytical derivation results is confirmed via Monte Carlo simulations.
It is demonstrated that increasing the number of elements of RIS can significantly improve the ABEP performance of the FD system over the half-duplex (HD) system. Furthermore, in the high SNR region, the ABEP performance of the FD system is better than that of the HD system.
\end{abstract}

\section {Introduction}
The fifth-generation (5G) wireless communications are expected to achieve spectral-efficient and energy-efficient transmissions, enabling a new vision for mobile communications that includes three main application scenarios: enhanced mobile broadband, ultra-reliability low-latency, and massive machine-type communications \cite{liu2021amul,wang2017joint}.
With the explosive growth of data traffic and the total number of connected devices, academia has begun to explore beyond 5G and sixth-generation (6G) technologies \cite{saad2020a}.
In response to emerging requirements and application scenarios, 6G communication technologies require a new communication paradigm, especially in the physical layer \cite{dang2020what}.
On this background, research has shifted its attention from both sides of the conventional transceiver to regulating the wireless channel, leading to the revolutionary reconfigurable intelligent surface (RIS) technique \cite{bash2021reconf}.
In addition, the introduction of RIS into other technological scenarios, such as RIS-assisted index modulation (IM) systems, has attracted widespread attention.

On the one hand, RIS is a two-dimensional planar array composed of a large number of reconfigurable metasurfaces \cite{wu2019inte}. Each element consists of a low-cost diode that can independently modulate the amplitude and phase of the incident signal.
Based on this, the RIS enables intelligent manipulation of the propagation environment to enhance received signal strength and achieve accurate beamforming, thus playing a key role in improving signal transmission reliability \cite{liu2023liu,li2013tow}.
It is worth noting that RIS is both flexible and easy to deploy, which can be deployed to enhance the desired signal or to cover blind areas \cite{renzo2020smart}.
In particular, deploying RIS near the transceiver can effectively alleviate the energy consumption burden.
Compared to relay, RIS does not require additional energy consumption for complex radio frequency (RF) processing, and the reflected signals do not introduce additional noise \cite{renzo2020recon,abd2020a}.
With the above characteristics, RIS provides a new approach and method for addressing problems in wireless communication research.

On the other hand, IM is a promising low-complexity modulation technique that leverages activation resources in space, time, frequency, or code domains for modulation information embedding, thus effectively enhancing the communication spectral efficiency \cite{mao2019novel,wen2017indexm}.
Spatial modulation (SM) is a promising technical solution based on conventional multiple-input multiple-output (MIMO) system architecture \cite{li2021single}. By equipping an RF chain at the transmitter side, additional information can be carried by a single antenna at each time slot, enabling a trade-off between energy efficiency and spectral efficiency \cite{zhu2022on}.
Besides, space shift keying (SSK) is a simplified version of SM that depends only on the antenna index to realize the information transmission and ignores the symbol field signal of SM.
In essence, both SM and SSK boost the modulation degrees of freedom by loading additional modulation into the antenna index.
However, in the millimeter-wave (mmWave) band, with the increase of transmission path loss, it is difficult to ensure the signal transmission quality with SM/SSK technology. For this reason, spatial scattering modulation (SSM) takes advantage of hybrid beams to effectively combat path loss \cite{ding2017spatial}. In particular, at each time slot, SSM employs the transmit beam to activate specific candidate scatterers in the channel instead of the transmit antenna \cite{zhu2023qua}.

\begin{table*}[t]
\centering
%\small
\caption{\small{Notations in this paper}}
\begin{tabular}{|c|l|c|l|}
\hline Notation & Definition & Notation & Definition \\
\hline$U_A$ & User A & $U_B$ & User B \\
\hline$N_t$ & Number of transmit antennas & $N_r$ & Number of receive antennas \\
\hline $P_A$ & Transmit power of $U_A$ & $P_B$ & Transmit power of $U_B$ \\
\hline$s_A$ & Transmit signal of $U_A$ & $s_{B}$ & Transmit signal of $U_B$ \\
\hline$l$ & Transmit antenna index of $U_B$ & $l'$ & Transmit antenna index of $U_A$ \\
\hline$\boldsymbol{g}_l^B$ & The channel from the $l$-th transmit antenna to RIS B & $\boldsymbol{g}_{l'}^A$ & The channel of the $l'$-th transmit antenna to RIS A \\
\hline$\mathbf{h}^A$ & The channel from RIS B to $U_A$ & $\mathbf{h}^B$ & The channel from RIS A to RIS B \\
\hline$\tilde {\mathbf{h}}^A$ & Self interference (SI) received at $U_A$ side & $\tilde {\mathbf{h}}^B$ & SI received at $U_B$ side  \\
\hline$h_{AA}$ & LI received at $U_A$ side & $h_{BB}$ & LI received at $U_B$ side \\
\hline$N_A$ & The variance of the noise received by $U_A$ & $N_B$ & The variance of the noise received by $U_B$ \\
\hline$\mathbb{C}^{m\times n}$ & Space of $m\times n$ matric & ${\rm diag (\cdot)}$ & Diagonal matrix operation \\
\hline$N$ & Number of the reflecting elements of RIS A and RIS B & $\mathcal{N}(\cdot,\cdot)$ & The Real Gaussian distribution \\
\hline $\mathcal{CN}(\cdot,\cdot)$ & The complex Gaussian distribution & $\Pr (\cdot)$ & Probability of happening of an event \\
\hline $\xi$ & The channel correlation coefficient & $|\cdot|$ & Absolute value operation \\
\hline $\Re\{\cdot\}$ & Real part operation & $\gamma_{th}$ & Threshold for signal outage \\
\hline$P_e$ & Conditional PEP & $\bar P_e$ & Unconditional PEP \\
\hline $E[\cdot]$ & The expectation operation & $Var[\cdot]$ & The variance operation \\
\hline $Q(\cdot)$ & The Q-function & $Q_\cdot(\cdot, \cdot)$ & The Marcum Q-function \\
\hline${\rm erf}(\cdot)$ & The Gaussian error function & $\cos(\cdot)$ & The cosine function \\
\hline$\sin(\cdot)$ & The sine function & $\exp(\cdot)$ &  The exponential function \\
\hline$f(\cdot)$ & PDF & $\sim$ & ``Distributed as"  \\
\hline$F(\cdot)$ & CDF &  $\frac{d}{dx}(\cdot)$ & Derivative operations with respect to $x$ \\
\hline$P_{\rm out}$ & Outage probability &  $\mathcal{T}$ &Throughput \\
\hline
\end{tabular}
\end{table*}
Recently, there has been an increasing interest in developing communication paradigms based on RIS and IM \cite{basar2020rec,canbilen2020ris,li2021space,yuan2021rece,jin2023ris,zhu2024on,zhu2023per,zhu2023rissk,singh2022ris,zhu2023RIS,on2023zhu,zhudouble23,zhu2022erm}.
To be specific, the authors in \cite{basar2020rec} proposed three RIS-IM schemes, i.e., IM for source transmit antennas, IM for RIS reflector regions, and IM for destination receive antennas.
It is shown that the system can achieve a relatively substantial average bit error probability (ABEP) performance improvement at a relatively low signal-to-noise ratio (SNR) compared to the system without RIS.
Then, \cite{basar2020rec} investigated the ABEP performance of the RIS-assisted received SM/SSK schemes with maximum likelihood (ML) and greedy detectors, respectively.
%\textcolor{blue}{In \cite{zhu2024on}, the authors investigated the ABEP performance of the RIS-SM scheme, where the SM technique is implemented at the transmitter.}
In \cite{zhu2024on}, the authors studied the ABEP performance of the RIS-SM scheme, where the SM technique is implemented at the transmitter.
Moreover, the author in \cite{canbilen2020ris} presented the transmit SSK for RIS systems and derived their performance based on the central limit theorem (CLT) in terms of ABEP.
After that, \cite{zhu2023rissk} and \cite{zhu2023per}  considered a more realistic communication scenario and analyzed the impact of channel estimation errors with respect to the ABEP performance of RIS-SSK systems.
Moreover, \cite{jin2023ris} proposed the RIS combined transceiver SSK reflection modulation scheme, where RIS can assist the transceiver SSK modulation and simultaneously introduce extra bits in the reflecting phase shift of the RF signal.
To improve the error and throughput performance,
\cite{li2021space} proposed RIS-SSK with passive beamforming and Alamouti space-time block coding schemes.
Additionally, \cite{yuan2021rece} investigated a RIS-assisted receive quadrature reflecting modulation, in which the whole RIS is divided into two halves to produce two orthogonal beams directed to the receive antenna.
In \cite{singh2022ris}, the RIS-SSK modulation and reflection phase modulation scheme, where the RIS embeds phase shift information during the reflection process.
In mmWave band, \cite{zhu2023RIS} and \cite{on2023zhu} investigated the RIS-assisted SSM system, where the RIS is placed closer to the transmitter to communicate with the line-of-sight (LoS) path and the receiver is farther away from the RIS to communicate with the none-LoS (NLoS) path.
In \cite{zhu2022erm}, the authors presented the RIS-double SSM (DSSM) scheme, in which the RIS is deployed in the middle of the channel, and the transmitter-to-RIS and RIS-to-receiver sub-channels are communicated with NLoS paths.
Next, \cite{zhudouble23} investigated the system performance of RIS-DSSM scheme in terms of the ABEP and ergodic capacity.

Full-duplex (FD) systems have been acknowledged as an important physical layer technology for simultaneous uplink and downlink communications with the same resource block. Compared to half-duplex (HD) systems with time--division or frequency-division duplex technology, FD systems can approximately double the spectral efficiency \cite{korp2016full}.
To improve the spectral efficiency of the RIS-SSK system, \cite{zhu20203fd} proposed the RIS-FD-SSK scheme and investigated the ABEP performance.
It is shown that as the spectral efficiency increases and the self-interference suppression capability improves, the performance of the FD system becomes more obvious than that of the HD system.
Further, \cite{zhu2023sic} divides a RIS uniformly into two parts used to assist uplink and downlink communications, respectively. In this way, the effect of residual self-interference on the ABEP performance of the system was investigated.
It is worth noting that the literature \cite{zhu20203fd} and \cite{zhu2023sic} mentioned above does not consider the transmission design with channel estimation error. Due to the unavoidable channel estimation error, if the estimated channel is naively considered the perfect channel, it will lead to a loss of system performance. As a result, robust transmission strategies must be designed for IRS-assisted wireless communication systems.
Against this background, this paper investigates the performance analysis of a RIS-assisted FD-SSK based system in the presence of imperfect CSI.
To the best of our knowledge, no works has been done on the RIS-based FD-SSK. It is more practical to study this system than previous work that considered perfect channels.
Overall, the contributions of this work are summarised as follows:

\begin{itemize}
\item We consider a RIS-assisted SSK communication system in which two FD users communicate with each other simultaneously. It is assumed that the physical distance between the two users is relatively far and the communication signal from the direct link is quite weak. Therefore, we consider that there is no direct link due to the presence of blockage. For this reason, we facilitate the signal transmission by equipping a dedicated RIS near each user. It is worth noting that a more realistic scenario of practical communication in the presence of channel estimation errors is considered.
\item At the receive antenna side, the transmit antenna is recovered employing the ML detection algorithm. The probability density function (PDF) of the combined channel in the case with optimal RIS phase shift is fitted with CLT. Based on the conditional pair error probability (PEP) and PDF, we derive integral expressions for the unconditional PEP. Since the exact closed-form expression is mathematically intractable, we apply the Gaussian-Chebyshev quadrature (GCQ) method to approximate the integral expression as the closed-form expression. Furthermore, the upper bound and asymptotic expression of ABEP are also provided.
\item Considering the optimal phase shift of RIS, the outage probability expression of the proposed RIS-FD-SSK scheme in the presence of channel estimation error is derived, where we derive the cumulative distribution function (CDF) expression of the proposed scheme taking advantage of the CLT-based composite channel. To gain more useful insights, we also provide the asymptotic outage probability expression and the lower bound of the system throughput for the considered systems.
\item We evaluate the error of the GCQ method used for the analytical derivation and the conditions for the applicability of the CLT. Afterwards, we validate the results of the analytical derivation through Monte Carlo simulations. Additionally, we investigate the effect of different parameters under the considered FD systems with respect to ABEP, outage probability, and throughput.
    Based on the obtained mathematical expressions, we find that
    the loop interference (LI) on the system performance can be suppressed effectively as the number of elements of RIS is relatively large. In addition, when the residual LI is relatively small or the number of units of RIS is large, the ABEP performance of the FD system is superior to that of the HD system. Furthermore, with both fixed and variable channel estimation errors, the FD system can achieve better ABEP performance than the HD system in the high SNR region.
\end{itemize}

The remainder of this paper is organized as follows. Section II describes the FD-SSK based RIS assisted system model and channel estimation error model as well as the detection algorithm. Section III derives the analytical expressions for ABEP, outage probability and throughput of the RIS-FD-SSK system under imperfect CSI. Simulation and analytical results confirming the analytical research are given in Section V. Section VI summarizes the whole paper.
Note that the Notations of this paper are summarized in Table I.

\section {System Model}
\begin{figure}[t]
  \centering
  \includegraphics[width=8.0cm]{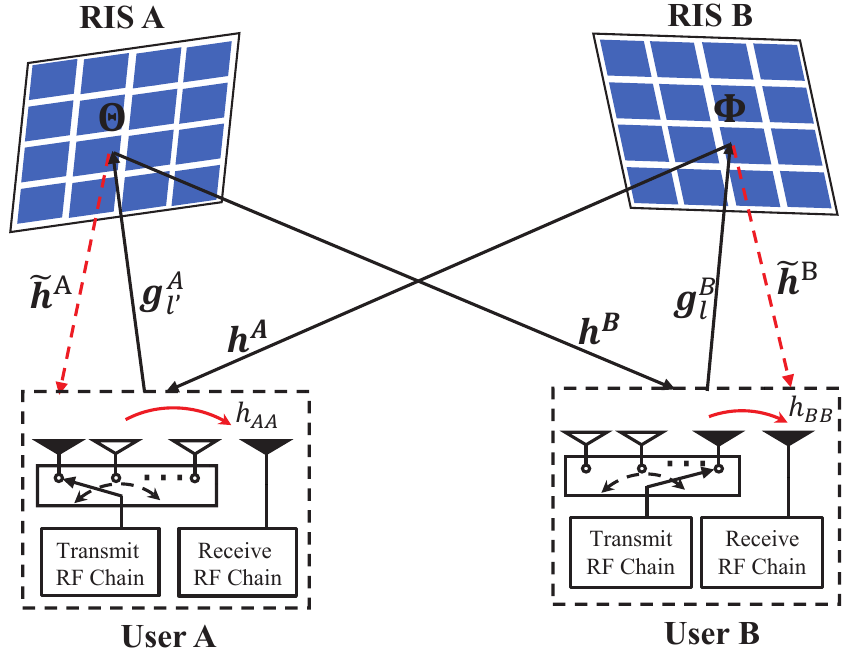}\\
  \caption{\small{System model.}}
  \label{system}
\end{figure}

Fig. \ref{system} depicts the system model of the proposed RIS-assisted FD-SSK system consisting of two users ($U_A$ and $U_B$) and two RISs (RIS A and RIS B).
We assume that $U_A$ and $U_B$ are far from each other and the direct link is blocked due to unfavorable environmental factors, in which the communication signal is very weak and can be negligible.
For this reason, we utilize RIS for assisted signal enhancement.
\cite{wu2019inte} demonstrated that deploying the RIS close to the transceiver results in better system performance than placing it in the middle of the channel. Inspired by this, we augment the signal emission by placing RIS A and RIS B near $U_A$ and $U_B$, respectively.

Since we employ the SSK technique on the transmit side of $U_A$ and $U_B$, only one transmit antenna is activated at each time slot based on the input bits.
Meanwhile, the phase shift matrix RIS can get the channel CSI based on the feedback link to align the reflection phase shift from the activated transmit antenna to the receive antenna.
Thus, the phase shift matrices $\boldsymbol{\Theta}$ of RIS A and $\boldsymbol{\Phi}$ of RIS B in Fig. \ref{system} would adjust the phase of the alignment channels from ${\boldsymbol{g}}_{l'}^A$ to $\mathbf{h}^B$ and from ${\boldsymbol{g}}_{l}^B$ to $\mathbf{h}^A$, respectively.
It is worth mentioning that in Fig. \ref{system} we do not consider the signal transmitted from the $U_A$ side activated antenna which is reflected by RIS B to reach $U_B$ and the signal transmitted from the $U_B$ side which is reflected by RIS A to reach $U_A$.
This is because the receive antenna path from the transmit antenna of $U_A$ reaching $U_B$ via RIS A and the receive antenna path from the transmit antenna of $U_B$ reaching $U_A$ via RIS A are asymmetric, and thus this part of the signal cannot be captured by the receive antenna of $U_A$.
Similarly, the receive antenna at the $U_B$ side is unable to capture the signal reflected from the transmitted signal of $U_A$ via RIS B.
Consequently, the received signal at $U_A$ and $U_B$ can be described as
\begin{equation}\label{eq1}
\begin{aligned}
{y}_A &\!\!=\! \!\sqrt{P_B}\mathbf{h}^A\boldsymbol\Phi{\boldsymbol{g}}_l^Bs_B\!\!+\!\! \sqrt{P_A} \tilde {\mathbf{h}}^A\boldsymbol\Theta {{\boldsymbol{g}}}_{l}^As_A \!\!+\!\!\sqrt{P_A}h_{AA} s_A\!+\! {n}_A,\\
\end{aligned}
\end{equation}
\begin{equation}\label{eq11}
\begin{aligned}
{y}_B &\!\!=\!\! \sqrt{P_A}\mathbf{h}^B\boldsymbol\Theta{\boldsymbol{g}}_{l'}^As_A\!\!+\!\! \sqrt{P_B}\tilde {\mathbf{h}}^B\boldsymbol\Phi{{\boldsymbol{g}}}_{l'}^Bs_B \!\!+\!\!\sqrt{P_B}h_{BB} s_B\!+\! {n}_B,
\end{aligned}
\end{equation}
where $\sqrt{P_B}\mathbf{h}^A\boldsymbol\Phi{\boldsymbol{g}}_l^Bs_B$ and $\sqrt{P_A}\mathbf{h}^B\boldsymbol\Theta{\boldsymbol{g}}_{l'}^As_A$ respectively represent the desired received signals of $U_A$ and $U_B$, where the reflection matrix of RIS A and RIS B can be expressed as $\boldsymbol\Phi = {\rm diag}(e^{j\phi_{1,l}},\ldots,e^{j\phi_{n,l}},\ldots,e^{j\phi_{N,l}})\in \mathbb{C}^{N\times N}$ and $\boldsymbol\Theta = {\rm diag}(e^{j\theta_{1,l'}},\ldots,e^{j\theta_{n,l'}},\ldots,e^{j\theta_{N,l'}})\in \mathbb{C}^{N\times N}$, respectively;
$ \sqrt{P_A} \tilde {\mathbf{h}}^A\boldsymbol\Theta {{\boldsymbol{g}}}_{l}^As_A$ and $\sqrt{P_B}\tilde {\mathbf{h}}^B\boldsymbol\Phi{{\boldsymbol{g}}}_{l'}^Bs_B$ respectively indicate the interference signals reflected by RIS A and RIS B;
$\sqrt{P_A}h_{AA} s_A$ and $\sqrt{P_B}h_{BB} s_B$ respectively denote the LI at the $U_A$ and $U_B$ sides;
$n_A$ and $n_B$ respectively denote the additive Gaussian white noise (AWGN) at the $U_A$ and $U_B$ sides, which are assumed to be circularly symmetric complex Gaussian (CSCG) distributed as $\mathcal{CN}(0,N_A)$ and $\mathcal{CN}(0,N_B)$.
Recall that RIS A and RIS B are located close to $U_A$ and $U_B$ sides, the channels of ${\boldsymbol{g}}_l^A$ and ${\boldsymbol{g}}_{l'}^B$ are relatively stable.
Therefore, the $U_A$-RIS A and $U_B$-RIS B channels are assumed as long-term stability, which can be estimated with high accuracy at low cost.
In contrast, RIS A and RIS B are farther away from the locations of $U_B$ and $U_A$, it is difficult to the inaccurately obtain the channel of $\mathbf{h}^B$ and $\mathbf{h}^A$.
Considering this, we assume that $U_A$-RIS A and $U_B$-RIS B channels can be perfected obtain, while $U_B$-RIS A and $U_A$-RIS B channel exist channel estimation errors.
Based on this, we have \cite{yang2022per}
\begin{equation}\label{eq2}
\mathbf{h}^A  = \xi\hat{\mathbf{h}}^A +\sqrt{1-\xi^2}\Delta{\mathbf{h}}^A,
\end{equation}
\begin{equation}\label{eq22}
\mathbf{h}^B  = \xi\hat{\mathbf{h}}^B +\sqrt{1-\xi^2}\Delta{\mathbf{h}}^B,
\end{equation}
where $\xi$ represents the correlation coefficient between the real channel and the estimated channel, $\hat {\mathbf{h}}^A$ and $\hat {\mathbf{h}}^B$ represent the channels obtained by using  pilot-assisted channel estimation approach such as least squares estimation or least mean square error estimation, and
$\Delta{\mathbf{h}}^A$ and $\Delta{\mathbf{h}}^B$ stand for the channel estimation errors, which are assumed to be independent of $\hat{\mathbf{h}}^A$ and $\hat{\mathbf{h}}^B$ and distributed as {$\mathcal{CN}(0,\sigma_e^2\mathbf{I}_{N\times N})$}.
Assuming that $\xi$ is known at the receiver side, two cases are considered in this paper. The first one is fixed $\sigma_e^2$: the value of the estimation error is constant for all SNR values to determine the pure effect of imperfect channel knowledge with respect to the error performance. The remaining scenario is the variable $\sigma_e^2$, that is, the value of the estimation error is adjusted based on the $\sigma_e^2=1/(\rho T)$ \cite{wang2011channel}, where $T$ denotes the number of pilot symbols used for training, and $\rho = \frac{P_A}{N_A}=\frac{P_B}{N_B}$ represents the SNR.
In views of (\ref{eq2}) and (\ref{eq22}), the received signal can be updated to
\begin{equation}\label{eq3}
\begin{aligned}
&y_A \!=\!  \sqrt{P_B\xi^2}\sum_{n=1}^N\hat{h}^A_ne^{j\phi_{n,l}}{g}_{n,l}^Bs_B\!+\!\sqrt{P_B(1\!-\!\xi^2)}\sum_{n=1}^N\Delta{{h}}^A_n e^{j\phi_{n,l}}\\&\times{g}_{n,l}^Bs_B\!+\! \sqrt{P_A} \sum_{n=1}^N\tilde {{h}}^A_ne^{j\theta_{n,l}} {{g}}_{n,l'}^As_A \!+\!\sqrt{P_A}h_{AA} s_A\!+ \!n_A,
\end{aligned}
\end{equation}
\begin{equation}\label{eq4}
\begin{aligned}
&y_B \!= \! \sqrt{P_A\xi^2}\sum_{n=1}^N\hat{{h}}_n^Be^{j\theta_{n,l}}{g}_{n,l'}^As_A\!+\!\sqrt{P_A(1\!-\!\xi^2)}\sum_{n=1}^N\Delta{h}^B_ne^{j\theta_{n,l}}\\&\times{g}_{n,l'}^As_A\!+ \! \sqrt{P_B}\sum_{n=1}^N\tilde {{h}}^B_ne^{j\phi_{n,l}}{{g}}_{n,l}^Bs_B \!+\!\sqrt{P_B}h_{BB} s_B\!+ \!n_B.
\end{aligned}
\end{equation}
In this respect, corresponding signal-interference-noise ratios (SINRs) are formulated at the bottom of the next page,
\begin{figure*}[b]
\hrulefill
\begin{equation}\label{eq5}
{\rm SINR_A} = \frac{{P_B\xi^2}\left|\sum_{n=1}^Na_nb_{n,l}e^{j(\phi_{n,l}-\psi_n-\vartheta_{n,l})}\right|^2}{{P_B(1-\xi^2)}\left|\sum_{n=1}^N\Delta{{h}}^A_nb_{n,l}e^{j(\phi_{n,l}-\vartheta_{n,l})}\right|^2 \!+\!P_A\left|\sum_{n=1}^N\tilde {{h}}^A_n \beta_{n,l'}e^{j(\theta_{n,l}-\varphi_{n,l'})}\right|^2\!+\!{P_A}|h_{AA}|^2 \!+\! N_A},
\end{equation}
\begin{equation}\label{eq6}
{\rm SINR_B} = \frac{{P_A\xi^2}\left|\sum_{n=1}^N \alpha_n\beta_{n,l'}e^{j(\theta_{n,l}-\delta_n-\varphi_{n,l'})}\right|^2}{{P_A(1\!-\!\xi^2)}\left|\sum_{n=1}^N\Delta{h}^B_n\beta_{n,l'}e^{j(\theta_{n,l}-\varphi_{n,l'})}\right|^2\!+ \!P_B\left|\sum_{n=1}^N\tilde {{h}}^B_nb_{n,l}e^{j(\phi_{n,l}-\vartheta_{n,l})}\right|^2\!+\!{P_B}|h_{BB}|^2 + N_B},
\end{equation}
\end{figure*}
where $\psi_n$, $\vartheta_{n,l}$, $\delta_n$, and $\varphi_{n,l'}$ denote the phases of $\hat h_n^A$, $g_{n,l}^B$, $\hat h_n^B$, and $g_{n,l'}^A$, respectively;
$a_n$, $b_{n,l}$, $\alpha_n$, and $\beta_{n,l'}$ means the amplitudes of $\hat h_n^A$, $g_{n,l}^B$, $\hat h_n^B$, and $g_{n,l'}^A$, respectively.
Since the desired signal is present in the numerator, the phase shift variables are adjusted to satisfy $\phi_{n,l}=\psi_n+\vartheta_{n,l}$ and $\theta_{n,l}=\delta_n+\varphi_{n,l'}$ to optimize the performance of its desired signal.
In this time, the corresponding SINRs are shown  at the bottom of the next page.
\begin{figure*}[b]
\begin{equation}\label{eq7}
{\rm SINR_A} = \frac{{P_B\xi^2}\left|\sum_{n=1}^Na_nb_{n,l}\right|^2}{{P_B(1-\xi^2)}\left|\sum_{n=1}^N\Delta{{h}}^A_nb_{n,l}e^{j\psi_n}\right|^2 +P_A\left|\sum_{n=1}^N\tilde {{h}}^A_n \beta_{n,l}e^{j\delta_n}\right|^2+{P_A}k_{AA}^2 + N_A},
\end{equation}
\begin{equation}\label{eq8}
{\rm SINR_B} = \frac{{P_A\xi^2}\left|\sum_{n=1}^N \alpha_n\beta_{n,l}\right|^2}{{P_A(1-\xi^2)}\left|\sum_{n=1}^N\Delta{h}^B_n\beta_{n,l}e^{j\delta_n}\right|^2+ P_B\left|\sum_{n=1}^N\tilde {{h}}^B_nb_{n,l}e^{j\psi_n}\right|^2+{P_B}k_{BB}^2 + N_B}.
\end{equation}
\end{figure*}
By applying interference cancellation techniques, the residual LI is still present in the system and distorts the desired signal. Thus, it is reasonable to model the residual LI as a noise source \cite{Ngu2021on}. It is worth mentioning that the $\sum_{n=1}^N\tilde {{h}}^A_n \beta_{n,l}e^{j\delta_n}$ and $\sum_{n=1}^N\tilde {{h}}^B_nb_{n,l}e^{j\psi_n}$ signals can be completely removed after interference cancellation in the digital domain \cite{Ngu2023inte,atapattu2020reconf}, the residual LI $I_{A}$ and $I_{B}$ at $U_A$ and $U_B$, respectively, can be expressed as $I_A\sim\mathcal{CN}(0,k_{AA}^2P_A)$ $I_B\sim\mathcal{CN}(0,k_{BB}^2P_B)$, where $k_{AA}^2$ and $k_{BB}^2$ stand for the residual SI level at the $U_A$ and $U_B$, respectively.

Considering this, (\ref{eq7}) and (\ref{eq8}) can be re-expressed as
\begin{equation}\label{eqs11}
{\rm SINR_A} = \frac{{P_B\xi^2}\left|\sum_{n=1}^Na_nb_{n,l}\right|^2}{{P_B(1\!-\!\xi^2)}\left|\sum_{n=1}^N\Delta{{h}}^A_nb_{n,l}e^{j\psi_n}\right|^2 \!\!+\!{P_A}k_{AA}^2 \!+\! N_A},
\end{equation}
\begin{equation}
{\rm SINR_B}\! =\! \frac{{P_A\xi^2}\left|\sum_{n=1}^N \alpha_n\beta_{n,l}\right|^2}{{P_A(1-\xi^2)}\left|\sum_{n=1}^N\Delta{h}^B_n\beta_{n,l}e^{j\delta_n}\right|^2+{P_B}k_{BB}^2 + N_B}.
\end{equation}
Recall that the SSK system only considers spatial domain signals, thus we can regard the symbol domain signals as $s_A=1$ and $s_B=1$.
In this way, received signals can be expressed as
\begin{equation}
y_A =  \sqrt{P_B\xi^2}\sum\nolimits_{n=1}^Na_nb_{n,l}+w_A,
\end{equation}
\begin{equation}
y_B =  \sqrt{P_A\xi^2}\sum\nolimits_{n=1}^N \alpha_n\beta_{n,l}+w_B,
\end{equation}
where
$w_A=\sqrt{P_B(1-\xi^2)}\sum_{n=1}^N\Delta{{h}}^A_nb_{n,l}e^{j\psi_n}+I_A+ n_A$
and
$w_B = \sqrt{P_A(1-\xi^2)}\sum_{n=1}^N\Delta{h}^B_n\beta_{n,l}e^{j\delta_n}+ I_B+ n_B$.

As a result, the received signals detected by ML detectors can be calculated as
\begin{equation}\label{mla}
[\hat l] = \arg\min\limits_{l\in\{1,\ldots,N_t\}}\left|y_A-\sqrt{P_B\xi^2}\sum\nolimits_{n=1}^Na_nb_{n,l}\right|^2,
\end{equation}

\begin{equation}
[\hat l'] = \arg\min\limits_{l'\in\{1,\ldots,N_t\}}\left|y_B-\sqrt{P_A\xi^2}\sum\nolimits_{n=1}^N \alpha_n\beta_{n,l'}\right|^2.
\end{equation}

\section{Performance Analysis}
In this section, our objective is to provide metrics that affect the proposed system via an analytical approach. Firstly, we utilize an ML detector to derive the PEP expression and provide the ABEP expression. Additionally, two metrics, outage probability and throughput, are provided.
It is worth noting the symmetry in the reception of signals by $U_A$ and $U_B$, and we only proceed with the subsequent derivation for the  $U_A$ side reception.
\subsection{Pairwise Error Probability (PEP)}

Based on (\ref{mla}), the conditional PEP can be given at the bottom of the next two pages,
\begin{figure*}[b]
\hrulefill
\begin{equation}\label{cpep1}
\begin{aligned}
P_e=&\Pr\left\{\left|y_A-\sqrt{P_B\xi^2}\sum\nolimits_{n=1}^Na_nb_{n,l}\right|^2>\left|y_A-\sqrt{P_B\xi^2}\sum\nolimits_{n=1}^Na_nb_{n,\hat l}e^{j(\vartheta_{n,l}-\vartheta_{n,\hat l})}\right|^2\right\}\\
=&\Pr\left\{\left|w_A\right|^2>\left|y_A\right|^2-2\Re\{
\sqrt{P_B\xi^2}\sum\nolimits_{n=1}^Na_nb_{n,\hat l}e^{j(\vartheta_{n,l}-\vartheta_{n,\hat l})}y_A
\}+{P_B\xi^2}\left|\sum\nolimits_{n=1}^Na_nb_{n,\hat l}e^{j(\vartheta_{n,l}-\vartheta_{n,\hat l})}\right|^2\right\}\\
=&\Pr\left\{\left|w_A\right|^2>\left|w_A\right|^2+\left|\sqrt{P_B\xi^2}\sum\nolimits_{n=1}^Na_nb_{n,l}\right|^2+2\Re\left\{\sqrt{P_B\xi^2}\sum\nolimits_{n=1}^Na_nb_{n,l}w_A\right\}\right.\\&\left.-2\Re\{
\sqrt{P_B\xi^2}\sum\nolimits_{n=1}^Na_nb_{n,\hat l}e^{j(\vartheta_{n,l}-\vartheta_{n,\hat l})}w_A
\}-2\left|\sqrt{P_B\xi^2}\sum\nolimits_{n=1}^Na_nb_{n,l}\right|^2\right.\\& \left.+{P_B\xi^2}\left|\sum\nolimits_{n=1}^Na_nb_{n,\hat l}e^{j(\vartheta_{n,l}-\vartheta_{n,\hat l})}\right|^2\right\}\\
=&\Pr\left\{-2\Re\left\{\sqrt{P_B\xi^2}\sum\nolimits_{n=1}^Na_n(b_{n,l}-b_{n,\hat l}e^{j(\vartheta_{n,l}-\vartheta_{n,\hat l})})w_A\right\}+\left|\sqrt{P_B\xi^2}\sum\nolimits_{n=1}^Na_nb_{n,l}\right|^2\right.\\
&\left.-\left|\sqrt{P_B\xi^2}\sum\nolimits_{n=1}^Na_nb_{n,\hat l}e^{j(\vartheta_{n,l}-\vartheta_{n,\hat l})}\right|^2>0\right\}\\
=&\Pr\left\{-\left|\sqrt{P_B\xi^2}\sum\nolimits_{n=1}^Na_n\left(b_{n,l}-b_{n,\hat l}e^{j(\vartheta_{n,l}-\vartheta_{n,\hat l})}\right)\right|^2\right.\\&\left.
-2\Re\left\{\sqrt{{P_B\xi^2}}\sum\nolimits_{n=1}^Na_n\left(b_{n,l}-b_{n,\hat l}e^{j(\vartheta_{n,l}-\vartheta_{n,\hat l})}\right)w_A\right\}
>0\right\}\\
=&\Pr\{F
>0\},
\end{aligned}
\end{equation}
\end{figure*}
where $F\sim\mathcal{N}(\mu_F,\sigma_F^2)$ with $\mu_F=-|\sqrt{P_B\xi^2}\sum\nolimits_{n=1}^Na_n(b_{n,l}-b_{n,\hat l}e^{j(\vartheta_{n,l}-\vartheta_{n,\hat l})})|^2$ and
$\sigma_F^2={2({P_B(1\!-\!\xi^2)}|\sum\nolimits_{n=1}^N\!\Delta{{h}}^A_nb_{n,l}e^{j\psi_n}|^2+\!{P_A}k_{AA}^2 \!\!+\!\! N_A)}$.
According to $\Pr(F>0)=Q(-\mu_F/\sigma_F)$, (\ref{cpep1}) can be further expressed as
\begin{small}
\begin{equation}\label{eqq1}
\begin{aligned}
P_e
&{=}Q\left(\sqrt{\frac{{P_B\xi^2}\left|\sum\limits_{n=1}^Na_n\left(b_{n,l}\!-\!b_{n,\hat l}e^{j(\vartheta_{n,l}\!-\!\vartheta_{n,\hat l})}\right)\right|^2}{2({P_B(1\!-\!\xi^2)}\!\left|\sum\limits_{n=1}^N\!\Delta{{h}}^A_nb_{n,l}e^{j\psi_n}\right|^2\! \!\!\!+\!{P_A}k_{AA}^2 \!\!+\!\! N_A)}}\right).
\end{aligned}
\end{equation}
\end{small}%
To facilitate subsequent analysis, let us make
\begin{equation}\label{equ}
\begin{aligned}
u &= \chi_l-\chi_{\hat l} =\sum_{n=1}^Na_nb_{n,l}-\sum_{n=1}^Na_nb_{n,\hat l}e^{j(\vartheta_{n,l}\!-\!\vartheta_{n,\hat l})},
\end{aligned}
\end{equation}
where $u$ is composed by summing $N$ mutually independent variables, the corresponding PDF form is very difficult to obtain directly.
Considering this, CLT is employed to conduct the solution of it. Firstly, we focus on the expectation and variance of $\sum_{n=1}^Na_nb_{n,l}$.
According to \cite{liu2022per}, the PDF of $|a_nb_{n,l}|$ can be calculated as
$
f_{|h_mg_m|}(x)=4xK_0(2x),
$
where $K_0(\cdot)$ is the second class of modified Bessel functions of order zero. Then, the mean and variance of {$|a_nb_{n,l}|$} can be characterized as
$
\mu_{|a_nb_{n,l}|}=\int_0^\infty 4x^2K_0(2x)dx=\frac{\pi}{4},
$
and
$
\sigma_{|a_nb_{n,l}|}^2=\int_0^\infty 4x^3K_0(2x)dx=\frac{16-\pi^2}{16},
$
respectively.
With the help of CLT, we can get
\begin{equation}\label{eqw25}
\mu_{\chi_l}=\frac{N\pi}{4}, \ \ \ {\sigma_{\chi_l}^2=\frac{N(16-\pi^2)}{16}}.
\end{equation}
Secondly, we deal with {$\sum_{n=1}^Na_nb_{n,\hat l}e^{j(\vartheta_{n,l}\!-\!\vartheta_{n,\hat l})}$}. It is worth mentioning that we resort to ${\bf Theorem 1}$ to obtain the expectation and variance of $b_{n,\hat l}e^{j(\vartheta_{n,l}-\vartheta_{n,\hat l})}$.
\begin{theorem}
The expectation and variance of $b_{n,\hat l}e^{j(\vartheta_{n,l}-\vartheta_{n,\hat l})}$ are respectively given by
\begin{equation}
\begin{aligned}
E[b_{n,\hat l}e^{j(\vartheta_{n,l}-\vartheta_{n,\hat l})}]&=0, \ \ Var[b_{n,\hat l}e^{j(\vartheta_{n,l}-\vartheta_{n,\hat l})}]&=1.
\end{aligned}
\end{equation}

Proof:
Without loss of generality, we make $x = \vartheta_{n,l}, y = \vartheta_{n,\hat l}$, and $z = \nu_n$.
Then, the PDF of $X$ and $Y$ can be expressed respectively as
$
f_X(x)=\frac{1}{2\pi}, f_Y(y)=\frac{1}{2\pi},  x,y\in[-\pi,\pi].
$
As a result, the PDF of the variable $z$ can be characterized as
$
f_Z(z)=\frac{d}{dz}F_Z(z),
$
where $F_Z(z)$ stands for the CDF of the variable $Z$, which can be formulated as
\begin{equation}\label{CDfz1}
\begin{aligned}
F_Z(z)&=\Pr\left(Z\leq z\right)=\Pr\left(X-Y\leq z\right)=\Pr\left(X\leq Y+z\right).
\end{aligned}
\end{equation}
Since $(\ref{CDfz1})$ contains two evevts $X$ and $Y$, it makes the analysis more complicated. To provide a more intuitive representation of event $Z \in[-2\pi,2\pi]$, we plot Fig. \ref{biao2}.

\begin{figure}[t]
 \centering
 \subfigure[Case 1: {$-2\pi\leq z\leq 0$.}]
 {
  \begin{minipage}[b]{0.22\textwidth}
   \centering
   \includegraphics[width=4cm]{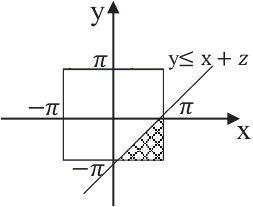}
  \end{minipage}
 }
 \subfigure[Case 2: {$0\leq z\leq 2\pi$.}]
    {
     \begin{minipage}[b]{0.22\textwidth}
      \centering
      \includegraphics[width=4cm]{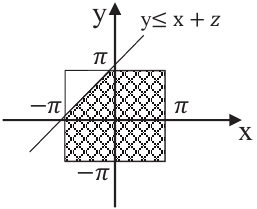}
     \end{minipage}
    }
\caption{\small{The integral region varies with the variable $z$.}}
\label{biao2}
\end{figure}

{\bf Case 1:} For $-2\pi\leq z\leq 0$, the shaded area in Fig. \ref{biao2}(a) indicates the variable interval of $x$ and $y$.
Here, the CDF of the $Z$ can be calculated as
$
F_Z(z)
=\int_{-\pi-z}^\pi\int_{-\pi}^{x+z} f(x,y)dydx,
$
where $f(x,y)$ represents the union PDF of variables $x$ and $y$.
Since $x$ and $y$ are two mutually independent variables, we have
$
F_Z(z)
=\int_{-\pi-z}^\pi\int_{-\pi}^{x+z} f(x)f(y)dydx.
$
After addressing the inner integral, we have
$
F_Z(z)
=\frac{1}{4\pi^2}\int_{-\pi-z}^\pi(x+z+\pi)dx.
$
By some simple algebraic operations, we can obtain
\begin{equation}\label{cdfz12}
F_Z(z)
=\frac{1}{4\pi^2}(3\pi^2+2z\pi+\frac{1}{2}z^2).
\end{equation}

{\bf Case 2:}
For $0\leq z\leq 2\pi$, the shaded area in Fig. \ref{biao2} stands for the variable interval of $x$ and $y$.
Since the area of the shaded part is somewhat complicated from the direction of the $x$-axis, we split it into a trapezoid from 0 to $\pi-z$ and a rectangle from $\pi-z$ to $\pi$. Accordingly, the CDF of $Z$ represented in Fig. 2(b) can be formulated as
$
F_Z(z)
=\int_{-\pi}^{\pi-z}\int_{-\pi}^{x+z}f(x,y)dydx+2\pi z.
$
Since $f(x,y)$ denotes the union PDF and $x$ and $y$ are independent of each other.
Therefore, we have
$
F_Z(z)
=\int_{-\pi}^{\pi-z}\int_{-\pi}^{x+z}f(x)f(y)dydx+2\pi z.
$
After tackling the inner integrals, we have
$
F_Z(z)
=\frac{1}{4\pi^2}\int_{-\pi}^{\pi-z}(x+z+\pi)dx+2\pi z.
$
After some manipulations, we have
\begin{equation}\label{cdfz14}
\begin{aligned}
F_Z(z)
&=\frac{1}{4\pi^2}\left(2\pi^2-\frac{1}{2}z^2\right)+2\pi z.
\end{aligned}
\end{equation}
Considering the (\ref{cdfz12}) and (\ref{cdfz14}) and taking the derivative operation, we have
\begin{equation}
f_Z(z)=
\begin{cases}
\frac{1}{4\pi^2}(2\pi+z), -2\pi<z<0\\
\frac{1}{4\pi^2}(2\pi-z), 0<z<2\pi.
\end{cases}
\end{equation}
On the basis of the error detection phase distribution $f_Z(z)$, the error detection symbol can be characterized as
\begin{equation}\label{mevai}
\begin{aligned}
E[b_{n,\hat l}e^{jz}]&=E[b_{n,\hat l}\cos{z}]+E[b_{n,\hat l}\sin{z}],\\
Var[b_{n,\hat l}e^{jz}]&=Var[b_{n,\hat l}\cos{z}]+Var[b_{n,\hat l}\sin{z}],
\end{aligned}
\end{equation}
where $E[b_{n,\hat l}]=\frac{\sqrt{\pi}}{2}$ and $Var[b_{n,\hat l}]=\frac{4-\pi}{4}$. Next, we aim at obtaining the mean and variance of $\cos z$ and $\sin z$, where the mean of of $\cos z$ and $\sin z$ can be respectively expressed as
$
E[\cos z]=
\!\!\int_{-2\pi}^0\!\!\frac{(2\pi+z)\cos z}{4\pi^2}dz+\!\int_0^{2\pi}\frac{(2\pi-z)\cos z}{4\pi^2}dz\!
=\!0
$ and
$
E[\sin z]=\!\!\int_{-2\pi}^0\!\!\frac{(2\pi+z)\sin z}{4\pi^2}dz+\!\int_0^{2\pi}\frac{(2\pi-z)\sin z}{4\pi^2}dz\!
=\!0.
$
Resort to $E[XY]=E[X]E[Y]$, we have
$
E[b_{n,\hat l}\cos{\nu_{n}}]=0$ and  $E[b_{n,\hat l}\sin{\nu_{n}}]=0.
$
To obtain the variance, $E[\cos^2 z]$ and $E[\sin^2 z]$ can be respectively derived as
\begin{equation}
\begin{aligned}
&E[\cos^2 z]=\int_{-2\pi}^0\frac{(2\pi+z)\cos^2 z}{4\pi^2}dz\!+\!\int_0^{2\pi}\frac{(2\pi-z)\cos^2 z}{4\pi^2}dz\\
&=\int_{-2\pi}^0\!\frac{(2\pi\!+\!z)(1\!+\!\cos 2z)}{8\pi^2}dz\!+\!\int_0^{2\pi}\!\frac{(2\pi-z)(1\!+\!\cos 2z)}{8\pi^2}dz\\
&=\frac{1}{8\pi^2}\left[\int_{-2\pi}^0(2\pi+z)dz+\int_0^{2\pi}(2\pi-z)dz\right]=\frac{1}{2},
\end{aligned}
\end{equation}
%and
\begin{equation}
\begin{aligned}
&E[\sin^2 z]=\int_{-2\pi}^0\frac{(2\pi+z)\sin^2 z}{4\pi^2}dz\!+\!\int_0^{2\pi}\frac{(2\pi-z)\sin^2 z}{4\pi^2}dz
\\
&=\!\int_{-2\pi}^0\!\!\frac{(2\pi\!+\!z)(1\!-\!\cos 2z)}{8\pi^2}dz\!+\!\int_0^{2\pi}\frac{(2\pi\!-\!z)(1\!-\!\cos 2z)}{8\pi^2}dz\\
&=\frac{1}{8\pi^2}\left[\int_{-2\pi}^0(2\pi+z)dz+\int_0^{2\pi}(2\pi-z)dz\right]=\frac{1}{2}.
\end{aligned}
\end{equation}
After that, the variance of $\sin z$ and $\cos z$ can be respectively given as
$
Var[\sin z]=E[\sin^2z]-E[\sin z]=\frac{1}{2}$ and
$Var[\cos z]=E[\cos^2z]-E[\cos z]=\frac{1}{2}.
$
According to $Var[XY]=E^2[X]Var[Y]+Var[X]E^2[Y]+Var[X]Var[Y]$, we have
$
E[b_{n,\hat l}\cos{\nu_{n}}]=0, Var[b_{n,\hat l}\cos{\nu_{n}}]
=\frac{1}{2},
E[b_{n,\hat l}\sin{\nu_{n}}]=0,  Var[b_{n,\hat l}\sin{\nu_{n}}]
=\frac{1}{2}.
$
On the basis of this, the (\ref{mevai}) can be represented as
\begin{equation}
\begin{aligned}
E[b_{n,\hat l}e^{jz}]&=0, \ \ Var[b_{n,\hat l}e^{jz}]&=1.
\end{aligned}
\end{equation}
By replacing $z$ with $\nu$, the proof is completed.
$\hfill\blacksquare$
\end{theorem}
Similar to the solution process of {\bf Theorem 1}, the expectation and variance of $a_nb_{n,\hat l}e^{j(\vartheta_{n,l}-\vartheta_{n,\hat l})}$ can be obtained as
$
E\left[a_nb_{n,\hat l}e^{j(\vartheta_{n,l}-\vartheta_{n,\hat l})}\right]=0$ and  $Var\left[a_nb_{n,\hat l}e^{j(\vartheta_{n,l}-\vartheta_{n,\hat l})}\right]=1.
$
By using CLT, we have
$
\mu_u = E\left[U\right]=\frac{\pi}{4}N$ and $\sigma^2_u = Var\left[U\right]=\frac{32-\pi^2}{16}N.
$
Accordingly, the PDF of $u$ can be given by
\begin{equation}
f_U(u) =\sqrt{\frac{8}{(32-\pi^2)N\pi}}\exp\left(-\frac{(4u-N\pi )^2}{64N-2N\pi^2}\right).
\end{equation}

To obtain the PDF of $U^2$ for subsequent PEP analysis of (\ref{eqq1}), let us define $X=U^2$, the CDF of numerator (\ref{eqq1}) can be formulated as
\begin{equation}
\begin{aligned}
F_X(x)&=\Pr\left(X\leq x\right)=\Pr\left(U^2\leq x\right)\\&=\Pr\left(-\sqrt{x}\leq U\leq \sqrt{x}\right)=\int_{-\sqrt{x}}^{\sqrt{x}}f_U(u)du.
\end{aligned}
\end{equation}
By calculating the derivative operation of the CDF of $X$, the corresponding PDF can be obtained as
\begin{equation}
\begin{aligned}
f_X(x)
%=&\frac{d}{dx}\int_{-\sqrt{x}}^{\sqrt{x}}f_U(u)du\\
=&\sqrt{\frac{2}{(32-\pi^2)N\pi{x}}}\left[\exp\left(-\frac{(4\sqrt{x}-N\pi )^2}{64N-2N\pi^2}\right)\right.\\&\left.+\exp\left(-\frac{(-4\sqrt{x}-N\pi )^2}{64N-2N\pi^2}\right)\right].
\end{aligned}
\end{equation}
After some calculation operations, we have
\begin{equation}
\begin{aligned}
f_X(x)=&\sqrt{\frac{2}{(32-\pi^2)N\pi  }}\exp\left(\frac{-\pi^2N^2}{64N-2N\pi^2}\right)\\&\times\frac{1}{\sqrt{x}}\exp\left(\frac{-8x}{32N-N\pi^2}\right)\\&\times
\left[\exp\left(\frac{4\pi\sqrt{x}}{32-\pi^2}\right)+\exp\left(\frac{-4\pi\sqrt{x}}{32-\pi^2}\right)\right].
\end{aligned}
\end{equation}
Let us make $\tau = \sqrt{\frac{2}{(32-\pi^2)N\pi x }}\exp\left(\frac{-\pi^2N^2}{64N-2N\pi^2}\right)$, the PDF can be given by
\begin{equation}\label{pdf1}
\begin{aligned}
f_X(x) =&\frac{\tau}{\sqrt{x}}\exp\left(\frac{-8x}{32N-N\pi^2}\right)\\&\times
\left[\exp\left(\frac{4\pi\sqrt{x}}{32-\pi^2}\right)+\exp\left(\frac{-4\pi\sqrt{x}}{32-\pi^2}\right)\right].
\end{aligned}
\end{equation}
For the $\left|\sum_{n=1}^N\Delta{{h}}^A_nb_{n,l}e^{j\psi_n}\right|^2$ term that lies in the denominator of (\ref{eqq1}).
According to the multiplication rules of independent random variables, we have $\Delta{{h}}^A_nb_{n,l}e^{j\psi_n}\sim\mathcal{CN}(0,\sigma_e^2)$.
By employing the CLT, we get $\sum_{n=1}^N\Delta{{h}}^A_nb_{n,l}e^{j\psi_n}\sim\mathcal{CN}(0,N\sigma_e^2)$.
In this respect, $\left|\sum_{n=1}^N\Delta{{h}}^A_nb_{n,l}e^{j\psi_n}\right|^2$
can be equivalently represented by power value $N\sigma_e^2$.
As a result, we can rewrite (\ref{eqq1}) as
\begin{equation}\label{eqq2}
\begin{aligned}
P_e
&=Q\left(\sqrt{\frac{{P_B\xi^2}x}{2({P_B(1-\xi^2)}N\sigma_e^2 +{P_A}k_{AA}^2 + N_A)}}\right).
\end{aligned}
\end{equation}
Combing (\ref{pdf1}) and (\ref{eqq2}), the PEP can be computed by
\begin{equation}\label{pe1}
\begin{aligned}
\bar P_e
=&\int_0^\infty \frac{\tau}{\sqrt{x}}\exp\left(\frac{-8x}{32N-N\pi^2}\right)\\ &\times
\left[\exp\left(\frac{4\pi\sqrt{x}}{32-\pi^2}\right)+\exp\left(\frac{-4\pi\sqrt{x}}{32-\pi^2}\right)\right]\\ &\times Q\left(\sqrt{\frac{{P_B\xi^2}x}{2({P_B(1-\xi^2)}N\sigma_e^2 +{P_A}k_{AA}^2 + N_A)}}\right)dx.
\end{aligned}
\end{equation}
According to $
Q(x)=\frac{1}{\pi}\int_0^{\frac{\pi}{2}}\exp\left(-\frac{x^2}{2\sin^2\theta}\right)d\theta
$ \cite{zhu20203fd}, then we exchange the order of integration.
Thus, the (\ref{pe1}) can be evaluated as
%\begin{equation}\label{pe2}
%\begin{aligned}
%\bar P_e
%=&\frac{\tau}{\pi}\int_0^\infty \int_0^{\frac{\pi}{2}}\frac{1}{\sqrt{x}}\exp\left(\frac{-8x}{32N-N\pi^2}-\frac{{P_B\xi^2}x}{4\sin^2\theta({P_B(1-\xi^2)}N\sigma_e^2 +{P_A}k_{AA}^2 + N_A)}\right)\\&\times
%\left[\exp\left(\frac{4\pi\sqrt{x}}{32-\pi^2}\right)+\exp\left(\frac{-4\pi\sqrt{x}}{32-\pi^2}\right)\right]d\theta dx.
%\end{aligned}
%\end{equation}
%Since it is very difficult to address (\ref{pe2}) directly, we exchange the order of integration as
\begin{equation}\label{peerx3}
\begin{aligned}
\bar P_e
=&\frac{\tau}{\pi} \int_0^{\frac{\pi}{2}}\int_0^\infty \frac{1}{\sqrt{x}}\exp\left(\frac{-8x}{32N-N\pi^2}\right.\\&\left.-\frac{{P_B\xi^2}x}{2\sin^2\theta({P_B(1-\xi^2)}N\sigma_e^2 +{P_A}k_{AA}^2 + N_A)}\right)\\&\times
\left[\exp\left(\frac{4\pi\sqrt{x}}{32-\pi^2}\right)+\exp\left(\frac{-4\pi\sqrt{x}}{32-\pi^2}\right)\right] dxd\theta.
\end{aligned}
\end{equation}
To simplify the representation, we let $\varsigma = \frac{2P_B\xi^2}{P_B(1-\xi^2)N\sigma_e^2+P_Ak_{AA}^2+N_A}$.
In views of this, (\ref{peerx3}) can be rewritten as
\begin{equation}\label{pe3}
\begin{aligned}
\bar P_e
=&\frac{\tau}{\pi}\int_0^{\frac{\pi}{2}}\int_0^\infty \frac{1}{\sqrt{x}}\exp\left(\frac{-8x}{32N-N\pi^2}-\frac{\varsigma x}{4\sin^2\theta}\right)\\&\times
\left[\exp\left(\frac{4\pi\sqrt{x}}{32-\pi^2}\right)+\exp\left(\frac{-4\pi\sqrt{x}}{32-\pi^2}\right)\right]dxd\theta.
\end{aligned}
\end{equation}
Without loss of generality, let us make $t = \sqrt{x}$.
Accordingly, (\ref{pe3}) can be recast as
\begin{equation}\label{barpxb}
\begin{aligned}
\bar P_e
=&\frac{2\tau}{\pi}\int_0^{\frac{\pi}{2}}\int_0^\infty \exp\left(-\frac{32(\sin^2\theta+N\varsigma)-N\varsigma\pi^2}{4(32N-N\pi^2)\sin^2\theta}t^2\right)\\&\times
\left[\exp\left(\frac{4\pi t}{32-\pi^2}\right)+\exp\left(\frac{-4\pi t}{32-\pi^2}\right)\right]dtd\theta.
\end{aligned}
\end{equation}
Let us define $\eta = \frac{32(\sin^2\theta+N\varsigma)-N\varsigma\pi^2}{4(32N-N\pi^2)\sin^2\theta}$, then (\ref{barpxb}) can be simplified as
\begin{equation}\label{barpb}
\begin{aligned}
\bar P_e
=&\frac{2\tau}{\pi}\int_0^{\frac{\pi}{2}}\int_0^\infty \exp\left(-{\eta t^2}\right)\\&\times
\left[\exp\left(\frac{4\pi t}{32-\pi^2}\right)+\exp\left(\frac{-4\pi t}{32-\pi^2}\right)\right]dtd\theta.
\end{aligned}
\end{equation}
To unlock the interior integrals of (\ref{barpb}), we appeal to the approach provided in \cite{jeff2007} as
\begin{equation}\label{tab207}
\int_0^\infty\exp\left(-\frac{x^2}{4\delta}-\gamma x\right)dx=\sqrt{\pi \delta}\exp(\delta\gamma^2)[1-{\rm erf}(\gamma\sqrt{\delta})].
\end{equation}
By substituting (\ref{tab207}) into (\ref{barpb}), the corresponding PEP can be expressed as
\begin{equation}\label{barpb1}
\begin{aligned}
&\bar P_e
=\frac{2\tau}{\pi}\int_0^{\frac{\pi}{2}}
\sqrt{\frac{\pi}{4\eta}}\exp\left(\frac{4\pi^2}{\eta(32-\pi^2)^2}\right)\\&\times\left[2-{\rm erf}\left(-\frac{2\pi}{(32-\pi^2)\sqrt{\eta}}\right)-{\rm erf}\left(\frac{2\pi}{(32-\pi^2)\sqrt{\eta}}\right)\right]
d\theta.
\end{aligned}
\end{equation}
Since ${\rm erf}(x)=\frac{2}{\sqrt{\pi}}\int_0^x \exp(-v^2)dv$ is an odd function, we have ${\rm erf}(x)+{\rm erf}(-x) = 0$.
Therefore, the (\ref{barpb1}) can be evaluated as
\begin{equation}\label{barpb2}
\begin{aligned}
\bar P_e
%=&\frac{2\tau}{\sqrt{\pi}}\int_0^{\frac{\pi}{2}}
%\sqrt{\frac{1}{\eta}}\exp\left(\frac{4\pi^2}{\eta(32-\pi^2)^2}\right)d\theta \\
=&\frac{2\tau}{\sqrt{\pi}}\int_0^{\frac{\pi}{2}}
\sqrt{\frac{4(32N-N\pi^2)\sin^2\theta}{32(\sin^2\theta+N\varsigma)-N\varsigma\pi^2}}\\&\exp\left(\frac{16N\pi^2\sin^2\theta}{(32(\sin^2\theta+N\varsigma)-N\varsigma\pi^2)(32-\pi^2)}\right)
d\theta.
\end{aligned}
\end{equation}
Note that (\ref{barpb2}) is an integral problem with respect to $\theta$ appearing in $(\ref{barpb2})$, which lies in the numerator and denominator, respectively.
It is very challenge to tackle this issue.
For this reason, we give the three solutions, which can be shown as follows.
\subsubsection{GCQ Method}
Subsequently, the substitution of variable $\theta = \frac{\pi}{4}\o+\frac{\pi}{4}$ (such that $d\theta=\frac{\pi}{4}d\o$; $\theta=0,\o=-1$; and $\theta=\frac{\pi}{2}, \o=1$) is used to reformulate (\ref{barpb2}) as
\begin{equation}\label{barpb4}
\begin{aligned}
&\bar P_e
=\frac{\tau\sqrt{\pi}}{2}\int_{-1}^1
\sqrt{\frac{4(32N-N\pi^2)\sin^2\left(\frac{\pi}{4}\o+\frac{\pi}{4}\right)}{32(\sin^2\left(\frac{\pi}{4}\o+\frac{\pi}{4}\right)+N\varsigma)-N\varsigma\pi^2}}\\&\times\exp\left(\frac{16N\pi^2\sin^2\left(\frac{\pi}{4}\o+\frac{\pi}{4}\right)}{(32(\sin^2\left(\frac{\pi}{4}\o+\frac{\pi}{4}\right)+N\varsigma)-N\varsigma\pi^2)(32-\pi^2)}\right)
d\o.
\end{aligned}
\end{equation}
By tackling the integral in (\ref{barpb4}) with the GCQ method, we get
\begin{equation}\label{barpb5}
\begin{aligned}
&\bar P_e
=\frac{\tau\sqrt{\pi^3}}{2Q}\sum_{q=1}^Q
\sqrt{\frac{4(1-\o_q^2)(3 2N-N\pi^2)\sin^2\left(\frac{\pi}{4}\o_q+\frac{\pi}{4}\right)}{32(\sin^2\left(\frac{\pi}{4}\o_q+\frac{\pi}{4}\right)+N\varsigma)-N\varsigma\pi^2}}\\
&\times\exp\left(\frac{16N\pi^2\sin^2\left(\frac{\pi}{4}\o_q+\frac{\pi}{4}\right)}{(32(\sin^2\left(\frac{\pi}{4}\o_q\!+\!\frac{\pi}{4}\right)\!+\!N\varsigma)\!-\!N\varsigma\pi^2)(32\!-\!\pi^2)}\right)\!+\!R_Q,
\end{aligned}
\end{equation}
where $Q$ represents the complexity accuracy trade-off parameter and
$R_Q$ denotes the error term that can be negligible at large values of $Q$.

\subsubsection{Upper bound}
According to $Q(x)\leq \frac{1}{12}\exp\left(-\frac{x^2}{2}\right)+\frac{1}{4}\exp\left(-\frac{2x^2}{3}\right)$, the (\ref{pe1}) can be approximated as

\begin{equation}\label{pej1}
\begin{aligned}
\bar P_e
\leq &\int_0^\infty \frac{\tau}{12\sqrt{x}}\exp\left(\frac{-8x}{32N-N\pi^2}\right)\\&\times
\left[\exp\left(\frac{4\pi\sqrt{x}}{32-\pi^2}\right)+\exp\left(\frac{-4\pi\sqrt{x}}{32-\pi^2}\right)\right] \\& \times \exp\left(-{\frac{{P_B\xi^2}x}{4({P_B(1-\xi^2)}N\sigma_e^2 +{P_A}k_{AA}^2 + N_A)}}\right)dx\\
&+\int_0^\infty \frac{\tau}{4\sqrt{x}}\exp\left(\frac{-8x}{32N-N\pi^2}\right)
\left[\exp\left(\frac{4\pi\sqrt{x}}{32-\pi^2}\right)\right.\\&\left.+\exp\left(\frac{-4\pi\sqrt{x}}{32-\pi^2}\right)\right] \\& \times \exp\left(-{\frac{{P_B\xi^2}x}{3({P_B(1-\xi^2)}N\sigma_e^2 +{P_A}k_{AA}^2 + N_A)}}\right)dx.
\end{aligned}
\end{equation}
Recall that $\varsigma = \frac{2P_B\xi^2}{P_B(1-\xi^2)N\sigma_e^2+P_Ak_{AA}^2+N_A}$, we can rewrite (\ref{pej1}) as
\begin{equation}\label{pex1}
\begin{aligned}
\bar P_e
%=&\int_0^\infty \frac{\tau}{12\sqrt{x}}\exp\left(\frac{-8x}{32N-N\pi^2}\right)
%\left[\exp\left(\frac{4\pi\sqrt{x}}{32-\pi^2}\right)+\exp\left(\frac{-4\pi\sqrt{x}}{32-\pi^2}\right)\right] \exp\left(-\frac{\varsigma x}{8}\right)dx\\
%&+\int_0^\infty \frac{\tau}{4\sqrt{x}}\exp\left(\frac{-8x}{32N-N\pi^2}\right)
%\left[\exp\left(\frac{4\pi\sqrt{x}}{32-\pi^2}\right)+\exp\left(\frac{-4\pi\sqrt{x}}{32-\pi^2}\right)\right] \exp\left(-\frac{\varsigma x}{6}\right)dx\\
\leq&\int_0^\infty \frac{\tau}{12\sqrt{x}}\exp\left(-\frac{64+\varsigma N(32-\pi^2)}{8N(32-\pi^2)}x\right)\\&\times
\left[\exp\left(\frac{4\pi\sqrt{x}}{32-\pi^2}\right)+\exp\left(\frac{-4\pi\sqrt{x}}{32-\pi^2}\right)\right] dx\\
&+\int_0^\infty \frac{\tau}{4\sqrt{x}}\exp\left(-\frac{48+\varsigma N(32-\pi^2)}{6N(32-\pi^2)}x\right)\\&\times
\left[\exp\left(\frac{4\pi\sqrt{x}}{32-\pi^2}\right)+\exp\left(\frac{-4\pi\sqrt{x}}{32-\pi^2}\right)\right] dx.
\end{aligned}
\end{equation}
By applying $t = \sqrt{x}$ for tractable analysis, (\ref{pex1}) can be rewritten as
\begin{equation}\label{pex2}
\begin{aligned}
\bar P_e
\leq&\frac{\tau}{6}\int_0^\infty \exp\left(-\frac{64+\varsigma N(32-\pi^2)}{8N(32-\pi^2)}t^2\right)\\&\times
\left[\exp\left(\frac{4\pi t}{32-\pi^2}\right)+\exp\left(\frac{-4\pi t}{32-\pi^2}\right)\right] dt\\
&+\frac{\tau}{2}\int_0^\infty \exp\left(-\frac{48+\varsigma N(32-\pi^2)}{6N(32-\pi^2)}t^2\right)\\&\times
\left[\exp\left(\frac{4\pi t}{32-\pi^2}\right)+\exp\left(\frac{-4\pi t}{32-\pi^2}\right)\right] dt.
\end{aligned}
\end{equation}
With the aid of the odd function properties of $\Phi(\cdot)$ and (\ref{tab207}), (\ref{pex2}) can be characterized as
\begin{equation}\label{pex3}
\begin{aligned}
\bar P_e
=&\frac{\tau}{6}\int_0^\infty \exp\left(-\frac{64+\varsigma N(32-\pi^2)}{8N(32-\pi^2)}t^2\right)\\&\times
\left[\exp\left(\frac{4\pi t}{32-\pi^2}\right)+\exp\left(\frac{-4\pi t}{32-\pi^2}\right)\right] dt\\
&+\frac{\tau}{2}\int_0^\infty \exp\left(-\frac{48+\varsigma N(32-\pi^2)}{6N(32-\pi^2)}t^2\right)\\&\times
\left[\exp\left(\frac{4\pi t}{32-\pi^2}\right)+\exp\left(\frac{-4\pi t}{32-\pi^2}\right)\right] dt\\
=&\frac{\tau}{3}\sqrt{ \frac{2\pi N(32-\pi^2)}{64+\varsigma N(32-\pi^2)}}\\&\times\exp\left(\frac{32N\pi^2}{64(32-\pi^2)+\varsigma N(32-\pi^2)^2}\right)\\
&+{\tau}\sqrt{\frac{3\pi N(32-\pi^2)}{2(48+\varsigma N(32-\pi^2))}}\\&\times\exp\left(\frac{48N\pi^2}{96(32-\pi^2)+2\varsigma N(32-\pi^2)^2}\right).
\end{aligned}
\end{equation}

%\begin{equation}\label{pej1}
%\begin{aligned}
%\bar P_e
%\leq &\int_0^\infty \frac{\tau}{2\sqrt{x}}\exp\left(\frac{-8x}{32N-N\pi^2}\right)
%\left[\exp\left(\frac{4\pi\sqrt{x}}{32-\pi^2}\right)+\exp\left(\frac{-4\pi\sqrt{x}}{32-\pi^2}\right)\right]\\
%&\times \exp\left(\frac{{-P_B\xi^2}x}{4({P_B(1-\xi^2)}N\sigma_e^2 +{P_A}k_{AA}^2 + N_A)}\right)dx.
%\end{aligned}
%\end{equation}
%Due to $\varsigma = \frac{2P_B\xi^2}{P_B(1-\xi^2)N\sigma_e^2+P_Ak_{AA}^2+N_A}$, (\ref{pej1}) can be reformulated as
%\begin{equation}\label{pex1}
%\begin{aligned}
%\bar P_e \leq&\int_0^\infty \frac{\tau}{2\sqrt{x}}\exp\left(-\frac{64\varsigma+32N-\pi^2N}{8\varsigma N(32-\pi^2)}x\right)
%\left[\exp\left(\frac{4\pi\sqrt{x}}{32-\pi^2}\right)+\exp\left(\frac{-4\pi\sqrt{x}}{32-\pi^2}\right)\right]dx.
%\end{aligned}
%\end{equation}
%By applying $t = \sqrt{x}$ for tractable analysis, (\ref{pex1}) can be rewritten as
%\begin{equation}\label{pex2}
%\begin{aligned}
%\bar P_e \leq &\int_0^\infty \tau\exp\left(-\frac{64\varsigma+32N-\pi^2N}{8\varsigma N(32-\pi^2)}t^2\right)
%\left[\exp\left(\frac{4\pi t}{32-\pi^2}\right)+\exp\left(\frac{-4\pi t}{32-\pi^2}\right)\right]dt.
%\end{aligned}
%\end{equation}
%With the aid of the odd function properties of $\Phi(\cdot)$ and (\ref{tab207}), (\ref{pex2}) can be characterized as
%\begin{equation}\label{barpb3}
%\begin{aligned}
%\bar P_e
%=&
%\sqrt{\frac{8\pi\tau^2\varsigma N(32-\pi^2)}{64\varsigma+32N-\pi^2N}}\exp\left(\frac{32\varsigma N\pi^2}{(64\varsigma+32N-\pi^2N)(32-\pi^2)}\right).
%\end{aligned}
%\end{equation}
\subsubsection{Asymptotic PEP}
To gain key insights into the proposed scheme, we derive the asymptotic PEP expression at high SNR values as
\begin{equation}\label{asy1d1}
\begin{aligned}
\lim\limits_{\rho\to\infty}\bar P_e
&=\lim\limits_{\rho\to\infty}\frac{2\tau}{\sqrt{\pi}}\int_0^{\frac{\pi}{2}}
\sqrt{\frac{4(32N-N\pi^2)\sin^2\theta}{32(\sin^2\theta+N\varsigma)-N\varsigma\pi^2}}\\&\times\exp\left(\frac{16N\pi^2\sin^2\theta}{(32(\sin^2\theta+N\varsigma)-N\varsigma\pi^2)(32-\pi^2)}\right)
d\theta.
\end{aligned}
\end{equation}
Since $\rho$ in (\ref{asy1d1}) is hidden in $\varsigma$, we take the limit operation on $\varsigma$ to obtain
\begin{equation}\label{asy1d2}
\begin{aligned}
\lim\limits_{\rho\to\infty}\varsigma
&= \frac{2P_B\xi^2}{P_B(1-\xi^2)N\sigma_e^2+P_Ak_{AA}^2+N_A}\\&= \frac{2\xi^2}{(1-\xi^2)N\sigma_e^2+k_{AA}^2 }.
\end{aligned}
\end{equation}
From this point, substituting (\ref{asy1d2}) into (\ref{asy1d1}), we get the corresponding asymptotic PEP.

\subsection{Average Bit Error Probability (ABEP)}
In this subsection, after obtaining PEP, the ABEP upper bound expression can be written as
\begin{equation}\label{abepx}
\begin{aligned}
{\rm ABEP} \leq \sum_{l=1}^{N_t}\sum_{l'=1,l'\neq l}^{N_t}\bar P_e N(l\to l'),
\end{aligned}
\end{equation}
where the term of $N(l\to l')$ stands for the Hamming distance of the corresponding PEP event.
\begin{remark}
The (\ref{abepx}) is the exact analytical ABEP  of the proposed scheme when $N_t=2$. In contrast, if $N_t >2$, (\ref{abepx}) is the tight upper bound analytical ABEP of the proposed scheme.
\end{remark}

\subsection{Outage Probability}
Apart from ABEP, the outage probability $P_{\rm out}$ is another metric for communication systems operating on fading channels. It is defined as the probability that the instantaneous ${\rm SINR_A}$ is below a specified threshold $\gamma_{th}$, i.e.
\begin{equation}\label{Ppout1}
\begin{aligned}
P_{\rm out} &=\Pr\left\{{\rm SINR_A}\leq \gamma_{th}\right\}\\&
\overset{(\ref{eqs11})}{=}\Pr\left\{\frac{{\rho\xi^2}\left|\chi_l\right|^2}{{\rho(1-\xi^2)}N\sigma_e^2 +{\rho}k_{AA}^2 + 1}\leq \gamma_{th}\right\}\\
&=\Pr\left\{\left|\chi_l\right|^2\leq \frac{{\rho(1-\xi^2)}N\sigma_e^2 +{\rho}k_{AA}^2 + 1}{\rho\xi^2} \gamma_{th}\right\}.
\end{aligned}
\end{equation}
Based on (\ref{eqw25}), it can be seen that $\chi_l$ follows the real Gaussian distribution. To make $z=\frac{{\rho(1-\xi^2)}N\sigma_e^2 +{\rho}k_{AA}^2 + 1}{\rho\xi^2} \gamma_{th}$ and $x=\chi_l$, (\ref{Ppout1}) can be rewritten as
\begin{equation}\label{Ppout2}
\begin{aligned}
P_{\rm out}
%&=\Pr\{-\sqrt{z}\leq \chi_l\leq \sqrt{z}\}\\
&=\frac{1}{\sqrt{2\pi\sigma_{\chi_l}^2}}\int_{-\sqrt{z}}^{\sqrt{z}} \exp\left({-\frac{(x-\mu_{\chi_l}^2)^2}{2\sigma_{\chi_l}^2}}\right)dx.
\end{aligned}
\end{equation}
To facilitate the subsequent analysis, (\ref{Ppout2}) is split into two indefinite integrals in the form that can be stated as
\begin{equation}\label{cdfx03}
\begin{aligned}
P_{\rm out}
=&\frac{\exp\left(-\frac{\mu^2_{\chi_l}}{2\sigma^2_{\chi_l}}\right)}{\sqrt{2\pi\sigma_{\chi_l}^2}}\int_{-\sqrt{z}}^\infty \exp\left({-\frac{x^2+2\mu_{\chi_l} x}{2\sigma_{\chi_l}^2}}\right)dx\\&-\frac{\exp\left(-\frac{\mu_{\chi_l}^2}{2\sigma_{\chi_l}^2}\right)}{\sqrt{2\pi\sigma_{\chi_l}^2}}\int_{\sqrt{z}}^\infty \exp\left({-\frac{x^2+2\mu_{\chi_l} x}{2\sigma_{\chi_l}^2}}\right)dx.
\end{aligned}
\end{equation}
For solving the integral in (\ref{cdfx03}), we appeal to \cite{jeff2007}
\begin{equation}\label{cdfx04}
\begin{aligned}
&\int_u^\infty\exp\left(-\frac{t^2}{4\delta}-\gamma t\right)dt\\&=\sqrt{\pi \delta}\exp(\delta\gamma^2)\left[1-{\rm erf}\left(\gamma\sqrt{\delta}+\frac{u}{2\sqrt{\delta}}\right)\right].
\end{aligned}
\end{equation}
Substituting (\ref{cdfx04}) into the (\ref{cdfx03}), we have
\begin{equation}
\begin{aligned}
P_{\rm out}
=&\frac{\exp\left(-\frac{\mu_{\chi_l}^2}{2\sigma_{\chi_l}^2}\right)}{\sqrt{2\pi\sigma_{\chi_l}^2}}\sqrt{\frac{\pi\sigma_{\chi_l}^2}{2}}\exp\left(\frac{\mu_{\chi_l}^2}{2\sigma_{\chi_l}^2}\right)\\&\times\left[1-{\rm erf}\left(\sqrt{\frac{z}{2\sigma_{\chi_l}^2}}-\sqrt{\frac{\mu_{\chi_l}^2}{2\sigma_{\chi_l}^2}}\right)\right]\\
&-\frac{\exp\left(-\frac{\mu_{\chi_l}^2}{2\sigma_{\chi_l}^2}\right)}{\sqrt{2\pi\sigma_{\chi_l}^2}}\sqrt{\frac{\pi\sigma_{\chi_l}^2}{2}}\exp\left(\frac{\mu_{\chi_l}^2}{2\sigma_{\chi_l}^2}\right)\\&\times\left[1-{\rm erf}\left(\sqrt{\frac{z}{2\sigma_{\chi_l}^2}}+\sqrt{\frac{\mu_{\chi_l}^2}{2\sigma_{\chi_l}^2}}\right)\right].
\end{aligned}
\end{equation}
After some manipulations, we have
\begin{equation}\label{cdft2}
\begin{aligned}
P_{\rm out}
=&1-\frac{1}{2}\left[{\rm erf}\left(\sqrt{\frac{z}{2\sigma_{\chi_l}^2}}-\sqrt{\frac{\mu_{\chi_l}^2}{2\sigma_{\chi_l}^2}}\right)\right.\\&\left.+{\rm erf}\left(\sqrt{\frac{z}{2\sigma_{\chi_l}^2}}+\sqrt{\frac{\mu_{\chi_l}^2}{2\sigma_{\chi_l}^2}}\right)\right].
\end{aligned}
\end{equation}
Since the Gaussian error function and the Q-function are related by ${{\rm erf}}(x)=2Q(\sqrt{2}x)$, we can reformulate (\ref{cdft2}) as
\begin{equation}\label{cdft3}
\begin{aligned}
P_{\rm out}
&=1\!-\!\left[Q\left(\sqrt{\frac{z}{\sigma_{\chi_l}^2}}\!-\!\sqrt{\frac{\mu_{\chi_l}^2}{\sigma_{\chi_l}^2}}\right)\!+\!Q\left(\sqrt{\frac{z}{\sigma_{\chi_l}^2}}\!+\!\sqrt{\frac{\mu_{\chi_l}^2}{\sigma_{\chi_l}^2}}\right)\right].
\end{aligned}
\end{equation}
For a threshold data rate of $R$ bps, the outage probability is given by
\begin{equation}\label{oplimi0}
P_{\rm out}=1\!-\!Q_{\frac{1}{2}}\left(\sqrt{\frac{\lambda}{\sigma_{\chi_l}^2}},\sqrt{\frac{(\rho(1\!-\!\xi^2)\sigma_e^2N\!+\!\rho k_{AA}^2\!+\!1)\gamma_{th}}{\rho\xi^2\sigma_{\chi_l}^2}}\right),
\end{equation}
where $\lambda = \mu_{\chi_l}^2$ and $\gamma_{th} = 2^{(R-\log_2N_t)}-1$.
\begin{figure*}[t]
 \centering
 \subfigure[{Error term analysis.}]
 {
  \begin{minipage}[b]{0.48\textwidth}
   \centering
   \includegraphics[width=8.7cm]{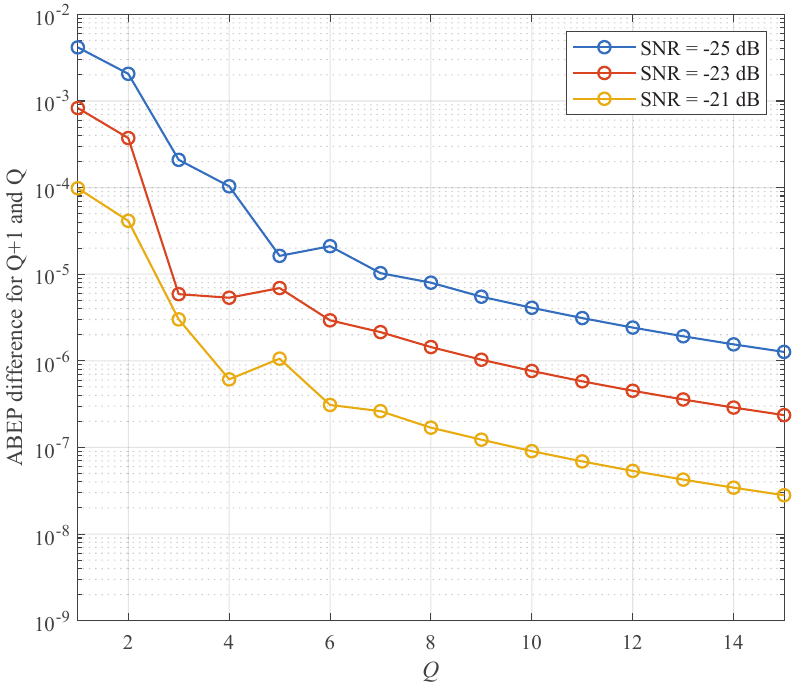}
  \end{minipage}
 }
 \subfigure[Convergence analysis.]
    {
     \begin{minipage}[b]{0.48\textwidth}
      \centering
      \includegraphics[width=8.7cm]{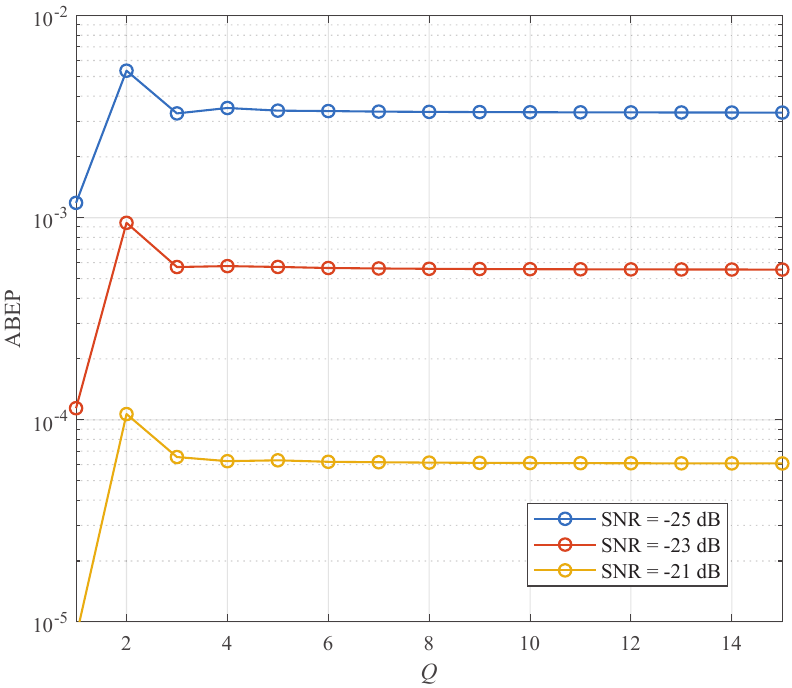}
     \end{minipage}
    }
\caption{\small{Verification of GCQ method.}}
\label{GCQ_cover1}
\end{figure*}
\subsubsection{Asymptotic Outage Probability}
In the high SNR region, the outage probability of (\ref{oplimi0}) can be evaluated as
\begin{equation}\label{oplimi1}
\lim\limits_{\rho \to \infty} P_{\rm out}=1-Q_{\frac{1}{2}}\left(\sqrt{\frac{\lambda}{\sigma_{\chi_l}^2}},\sqrt{\frac{(1-\xi^2)\sigma_e^2N+k_{AA}^2}{\xi^2\sigma_{\chi_l}^2}\gamma_{th}}\right).
\end{equation}
Recall that $\xi=\frac{1}{\sqrt{1+\sigma_e^2}}$, then (\ref{oplimi1}) can be recast as
\begin{equation}
\lim\limits_{\rho \to \infty} P_{\rm out}=1-Q_{\frac{1}{2}}\left(\sqrt{\frac{\lambda}{\sigma_{\chi_l}^2}},\sqrt{\frac{\sigma_e^4N+(1+\sigma_e^2)k_{AA}^2}{\sigma_{\chi_l}^2}\gamma_{th}}\right).
\end{equation}

\begin{remark}
By taking the limit of $\rho$, it can be observed that the result of the asymptotic outage probability is independent of the $\rho$. This indicates that the asymptotic outage probability is a constant value in the high SNR region.
\end{remark}

\subsection{Throughput}
In this subsection, the throughput can be expressed in terms of the number of correctly detected bits. In this regard, the throughput can be characterized as \cite{tse2005book}
\begin{equation}\label{throughput}
\mathcal{T}=\frac{(1-{\rm ABEP})}{T_s}\log_2(N_t),
\end{equation}
where $(1-{\rm ABEP})$ stands for the probability that the received bits are correctly detected during the transmission time slot $T_s$.

\begin{figure*}[t]
 \centering
 \subfigure[CLT results.]
 {
  \begin{minipage}[b]{0.48\textwidth}
   \centering
   \includegraphics[width=8.7cm]{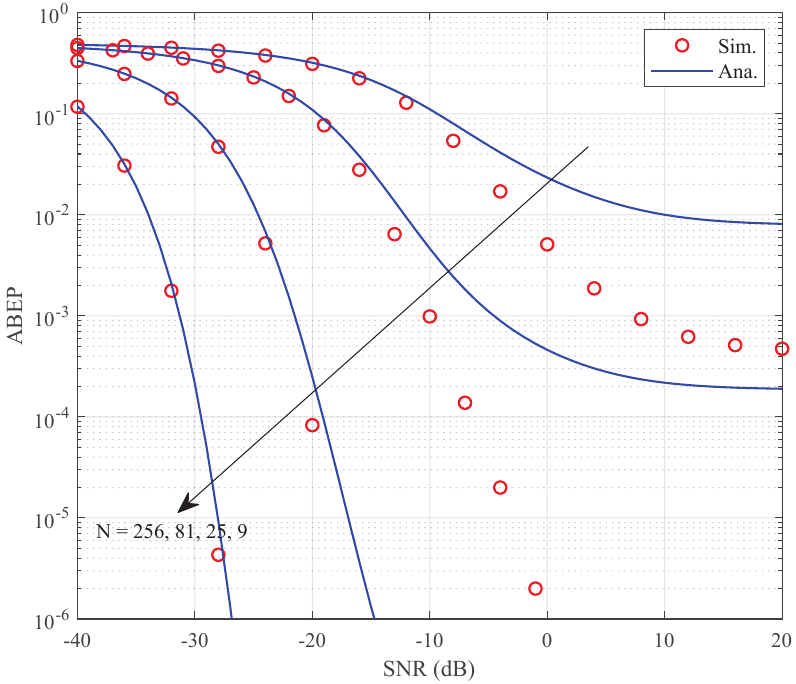}
  \end{minipage}
 }
 \subfigure[{Theoretical results.}]
    {
     \begin{minipage}[b]{0.48\textwidth}
      \centering
      \includegraphics[width=8.7cm]{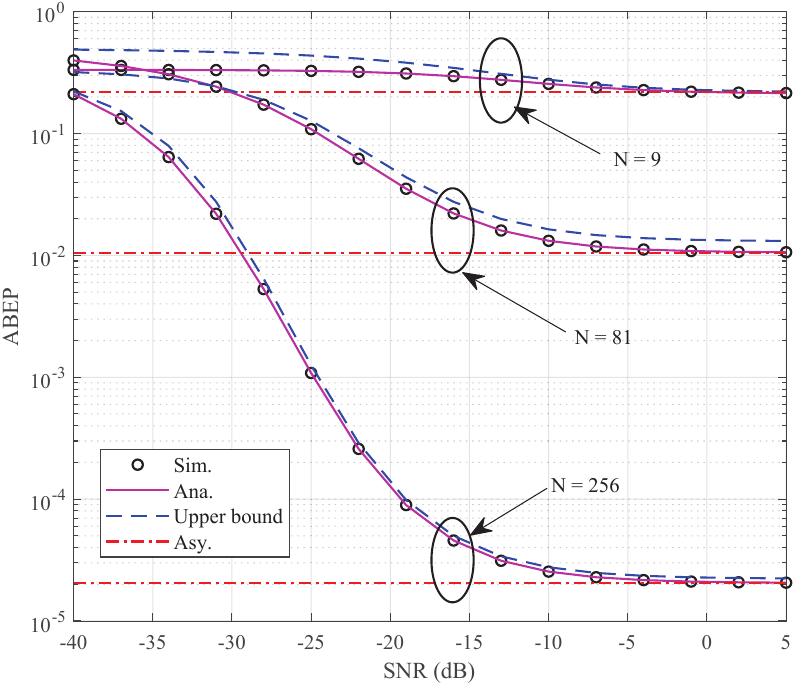}
     \end{minipage}
    }
\caption{\small{{Verification of correctness of analytical derivation.}}}
\label{CLT_ber}
\end{figure*}
\begin{figure*}[t]
 \centering
 \subfigure[{ $N=25$}]
 {
  \begin{minipage}[b]{0.48\textwidth}
   \centering
   \includegraphics[width=8.7cm]{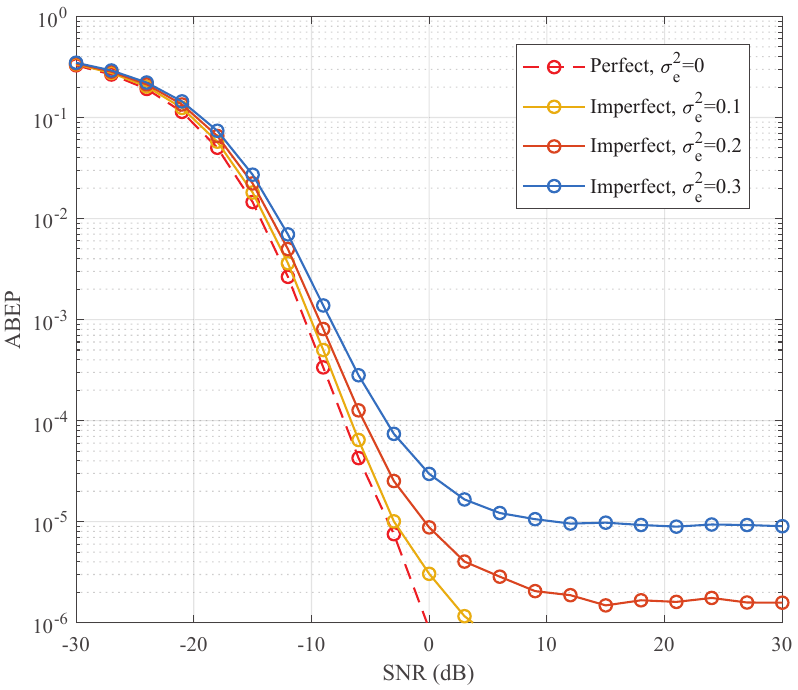}
  \end{minipage}
 }
 \subfigure[{ $N=256$}]
    {
     \begin{minipage}[b]{0.48\textwidth}
      \centering
      \includegraphics[width=8.7cm]{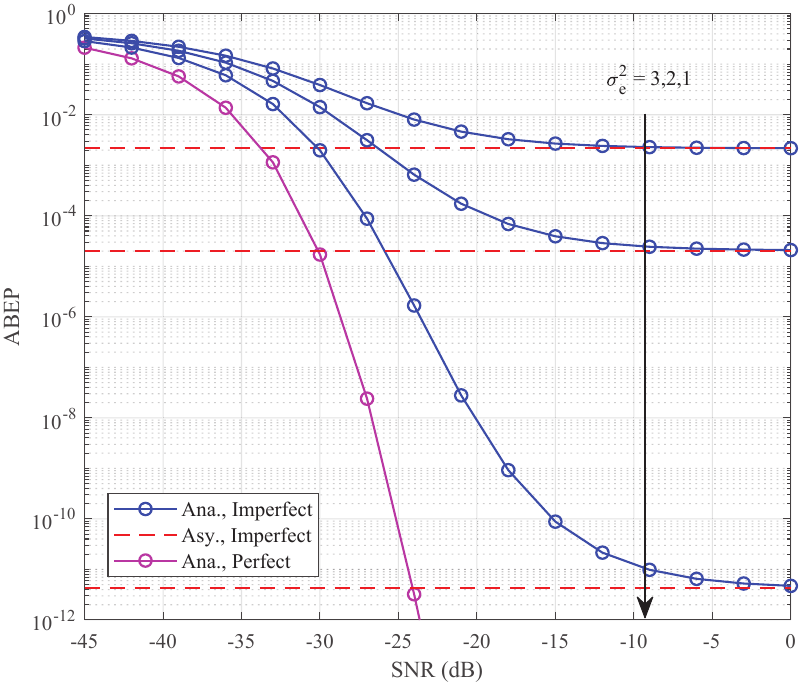}
     \end{minipage}
    }
\caption{\small{Impact of fixed $\sigma_e^2$ on the proposed systems.}}
\label{ImperfectCSI}
\end{figure*}
\begin{figure*}[t]
 \centering
 \subfigure[{ $N=25$}]
 {
  \begin{minipage}[b]{0.48\textwidth}
   \centering
   \includegraphics[width=8.7cm]{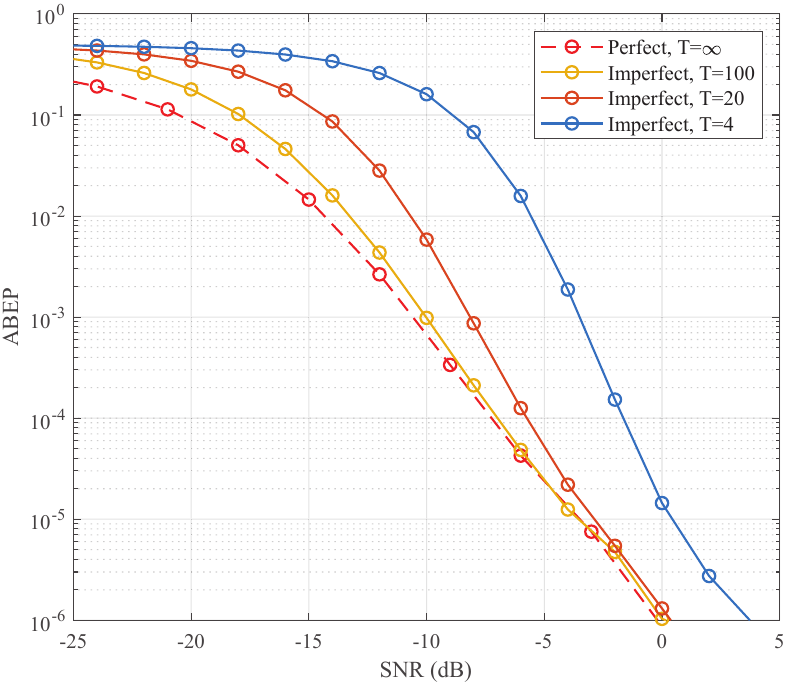}
  \end{minipage}
 }
 \subfigure[{ $N=256$}]
    {
     \begin{minipage}[b]{0.48\textwidth}
      \centering
      \includegraphics[width=8.7cm]{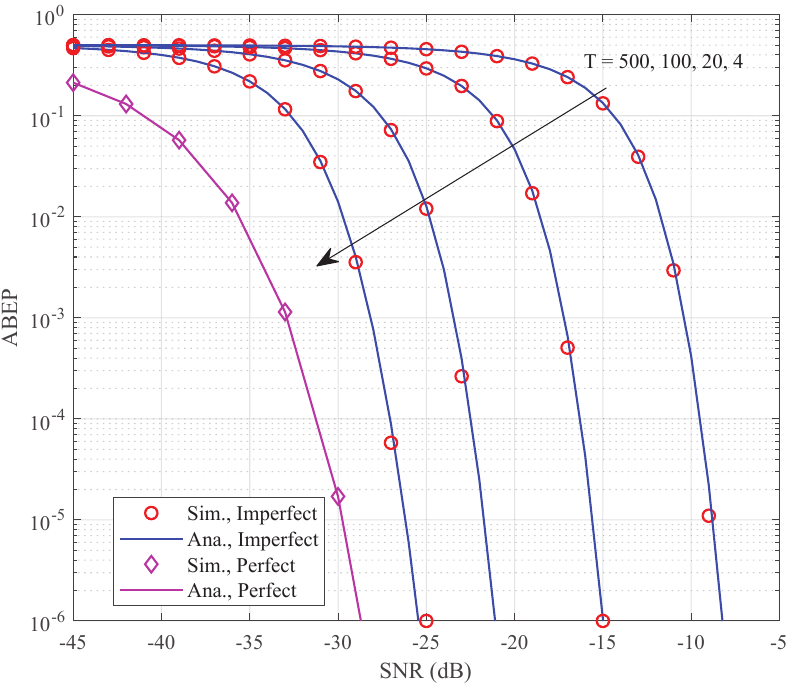}
     \end{minipage}
    }
\caption{\small{Impact of variable $\sigma_e^2=\frac{1}{T\rho }$ on the proposed systems.}}
\label{Imperfect_versmal}
\end{figure*}

\section{Simulation and Analytical Results}
In this section, Monte Carlo simulations are applied to confirm the correctness of the derived results. Moreover, the effect of all system parameters on the ABEP, outage probability, and throughput are investigated.
Additionally, the ABEP of RIS-assisted FD-SSK with perfect CSI and HD-SSK systems are provided for comparison. In all scenarios, the system parameters are set as follows: each simulation value is generated by a $10^6$ channel realization.
Unless otherwise specified, the transmitting antenna $N_t$ is set to 2.

\subsection{Verification}
Fig. \ref{GCQ_cover1}(a) illustrates the ABEP difference obtained for GCQ error term  versus the complexity-accuracy trade-off parameter $Q$ for different values of SNR.
The parameters are set $k_{AA}^2$, $N$,and $\sigma_e^2$ as $0.1$, $100$, and 0.1, respectively. In particular, we consider different cases with SNRs of $-25$ dB, $-23$ dB, and $-21$ dB.
It is clearly shown that the ABEP value of the difference between two adjacent values of $Q$ decreases as the SNR increases.
As $Q$ is greater than or equal to 6, the magnitude of change in the ABEP difference tends to stabilize as $Q$ increases, while at $Q$ less than 6, the magnitude of change in the ABEP difference is less stable. It is worth mentioning that when $Q$ is 3, the ABEP difference is less than $10^{-3}$, which can be almost negligible.

To more intuitively measure the impact of $Q$ value versus ABEP value, we plot the Fig. \ref{GCQ_cover1}(b), in which the parameter settings are consistent with Fig. \ref{GCQ_cover1}(a).
It can be clearly seen that, when $Q$ takes a value of 6 and beyond, ABEP tends to be stable. If $Q$ is 3, the difference between the ABEP value and $Q=6$ is quite small and almost negligible.
As a result, when the requirements for ABEP accuracy are not very high, $Q=3$ can reduce the computational complexity compared to $Q=6$.

\begin{figure*}[t]
 \centering
 \subfigure[{ $N=9$}]
 {
  \begin{minipage}[b]{0.48\textwidth}
   \centering
   \includegraphics[width=8.7cm]{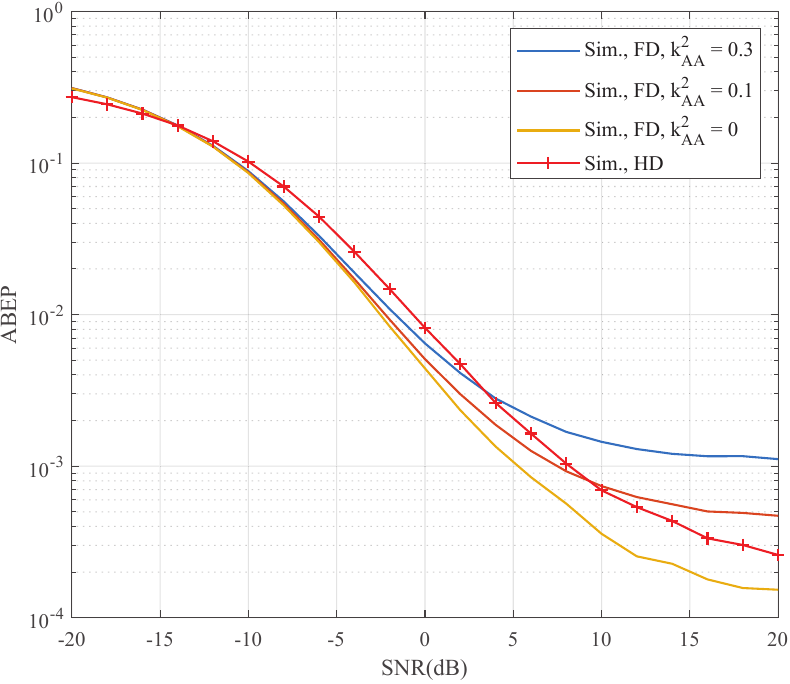}
  \end{minipage}
 }
 \subfigure[{ $N=16$}]
    {
     \begin{minipage}[b]{0.48\textwidth}
      \centering
      \includegraphics[width=8.7cm]{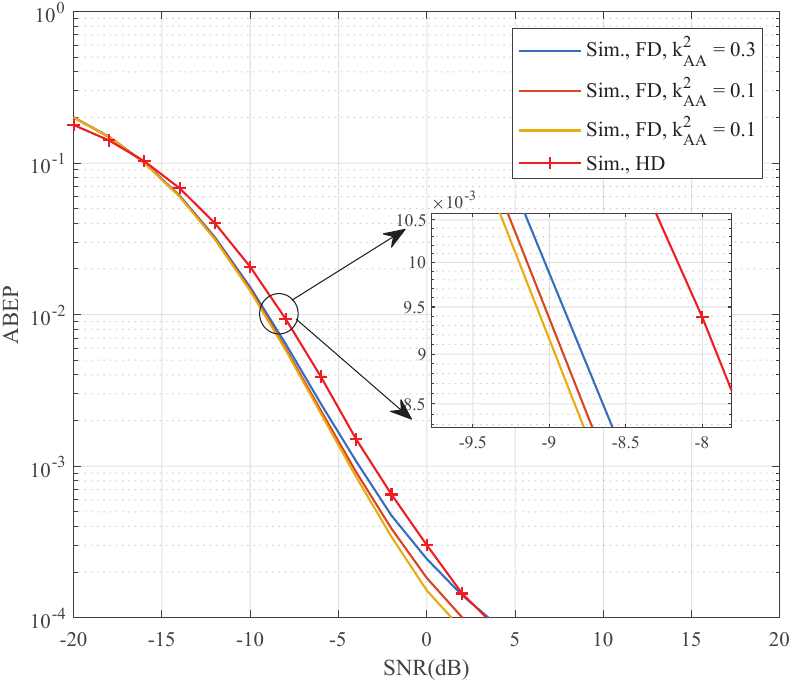}
     \end{minipage}
    }
\caption{\small{{Impact of residual LI on RIS-FD-SSK vs. RIS-HD-SSK systems}.}}
\label{LI_16}
\end{figure*}

Fig. \ref{CLT_ber}(a) presents the relationship between ABEP and SNR with the number of RIS elements, where $k_{AA}^2=0.1$ and $\sigma_e^2=0.1$. It can be observed by Fig. \ref{CLT_ber}(a) that ABEP decreases significantly with the number of RIS reflecting elements from 9 to 256, which is due to the fact that the larger the RIS is, the more energy can be reached at the receive antenna.
Besides, the difference between the simulation results and the analytical curves is large at small values of $N$, but as $N$ increases, the gap between the two decreases until they match perfectly.
This is because when the value of $N$ is relatively small, the CLT application conditions do not hold, On the contrary, when the value of $N$ is relatively large, the distribution of the combined channel can be equivalent to a Gaussian distribution.

In Fig. \ref{CLT_ber}(b), we investigate the relationship between the simulation, analytical, upper bound, and asymptotic results in terms of ABEP for the RIS-FD-SSK scheme with imperfect CSI, where $k_{AA} ^2=0.1$ and $\sigma_e^2=2$.
In particular, the upper bound and asymptotic curves of ABEP are plotted through (\ref{barpb2}) and (\ref{asy1d1}), respectively.
As expected, the simulation and analysis results match each other perfectly over the entire SNR region.
Moreover, it can be observed that there exists a small gap between the upper bound of ABEP and the exact analytical result, which suggests that (\ref{pex3}) is the relatively tight upper bound of ABEP across the entire SNRs.
Moreover, as the SNR gradually increases, the analytical results of ABEP tend to be a fixed constant and are in high agreement with the asymptotic ABEP results, which verifies the correctness of the derivation of the asymptotic ABEP expression.

\subsection{System Performance Under Various Parameters}

In Fig. \ref{ImperfectCSI}(a), we show the effect of the fixed channel estimation error parameter $\sigma_e^2$ on ABEP when the number of RIS reflection elements $N=25$. Recall that the analytical ABEP is obtained based on CLT, which means that the error of the analytical result is large as the number of reflection elements is small. Thus, we illustrate this situation via simulation, where we set $k_{AA}^2=0.1$.
We can observe from Fig. \ref{ImperfectCSI}(a) that ABEP decreases and approaches the perfect CSI case as the fixed channel estimation error $\sigma_e^2$ becomes smaller.
However, in the case of imperfect CSI, as the SNR increases, ABEP is subject to error floor. Furthermore, when the error is smaller, the error floor is approximately lower. This phenomenon implies that the factor dominating ABEP is no longer the SNR but the estimation error variance.

Fig. \ref{ImperfectCSI}(b) compares the ABEP for two scenarios with different channel estimation error variances and perfect CSI. The results show that the imperfections of the CSI with channel estimation errors have a significant impact on the ABEP of the proposed system.
For example, at SNR = 0 dB, the channel estimation variance $\sigma_e^2$ of 3, 2, and 1 corresponds to ABEPs of about $10^{-3}, 10^{-5}$, and $1.3\times 10^{-12}$, respectively.
In addition, in the case of imperfect CSI, ABEP performance gradually slows down as SNR increases and eventually approaches a constant. Furthermore, the asymptotic ABEP matches perfectly with the ABEP in the high SNR region. At this time, since the ABEP in the high SNR region is close to the constant lower limit, the diversity gain of the system is zero.

\begin{figure*}[t]
 \centering
 \subfigure[{ Fixed $\sigma_e^2$}]
 {
  \begin{minipage}[b]{0.48\textwidth}
   \centering
   \includegraphics[width=8.7cm]{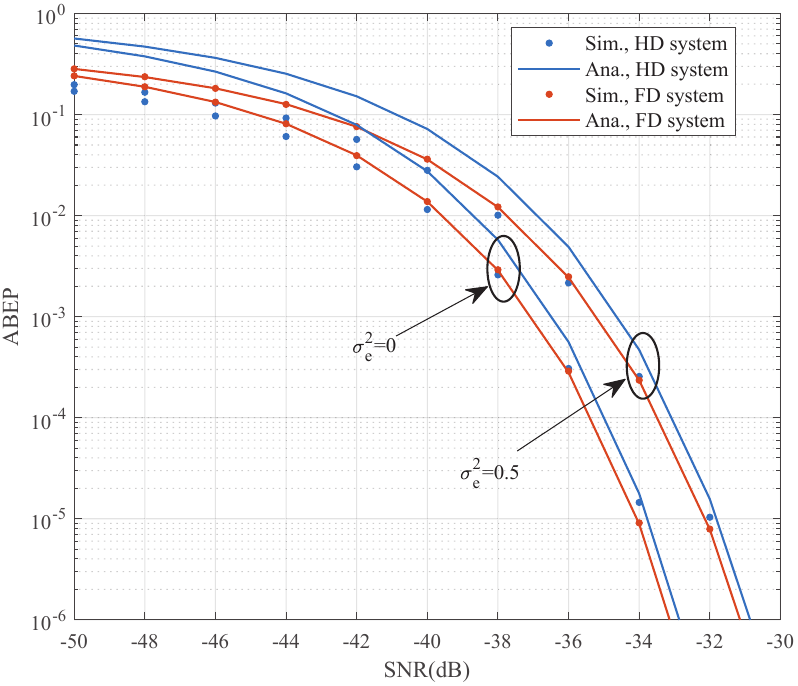}
  \end{minipage}
 }
 \subfigure[{ Variable $\sigma_e^2=\frac{1}{T\rho }$}]
    {
     \begin{minipage}[b]{0.48\textwidth}
      \centering
      \includegraphics[width=8.7cm]{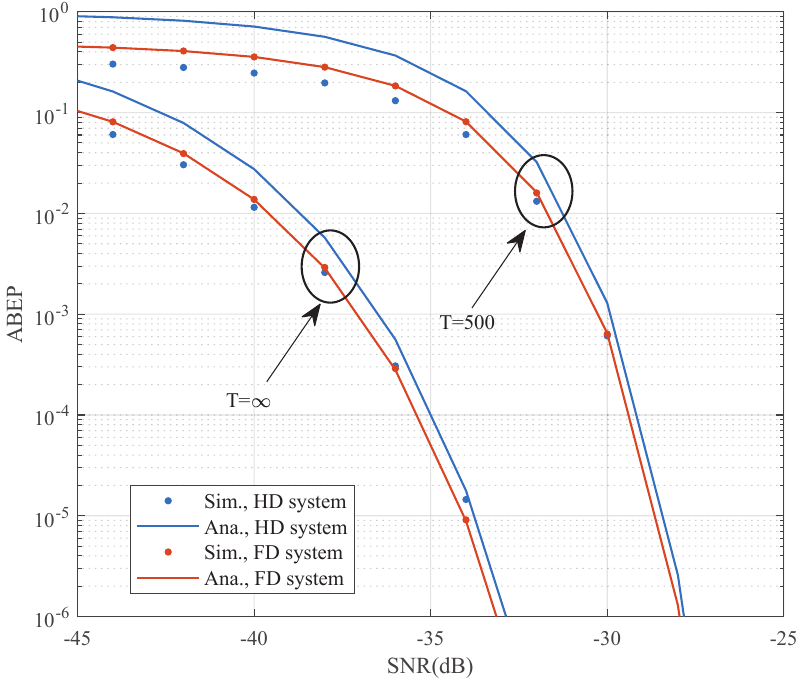}
     \end{minipage}
    }
\caption{\small{{Impact of imperfect CSI on RIS-FD-SSK vs. RIS-HD-SSK systems.}}}
\label{contras_jin_high}
\end{figure*}

Fig. \ref{Imperfect_versmal}(a) illustrates ABEP versus SNR with respect to the variable channel estimation variance $\sigma_e^2$ as the number of RIS reflection elements is 25, where the LI parameter is set as the $k_{AA}^2=0.1$.
According to Fig. \ref{CLT_ber}, it can be seen that
the error between the analytical results based on CLT and the simulation results is large
for lower values of $N$.
Considering this, the curves in Fig. \ref{Imperfect_versmal}(a) are obtained by Monte Carlo simulation.
More specifically, we can observe from Fig. \ref{Imperfect_versmal}(a) that the imperfect CSI case effectively reduces the ABEP value as the number of pilots $T$ increases, because the higher the number of pilots, the higher the accuracy of channel estimation. Again, as the SNR increases, the ABEP in the imperfect CSI case gradually converges to the perfect CSI, which is caused by the unit of increase of the SNR in dB.

Fig. \ref{Imperfect_versmal}(b), we plot the ABEP versus SNR in different variable channel estimation variances $\sigma_e^2$.
It is worth mentioning that, except for the number of reflection units $N$ and the number of pilots $T$, the rest of the parameter settings are consistent with Fig. \ref{Imperfect_versmal}(a).
As expected, when $N=256$, the simulation results match the analytical results highly, which is due to the fact that each element of the RIS is independent of each other and the composite channel obeys a Gaussian distribution.
Compared with Fig. \ref{Imperfect_versmal}(a), Fig. \ref{Imperfect_versmal}(b) requires more pilots to approach the perfect CSI situation. This is because as the number of RIS reflection elements increases, the required pilot overhead increases.

\begin{figure}[t]
\centering
\includegraphics[width=8.7cm]{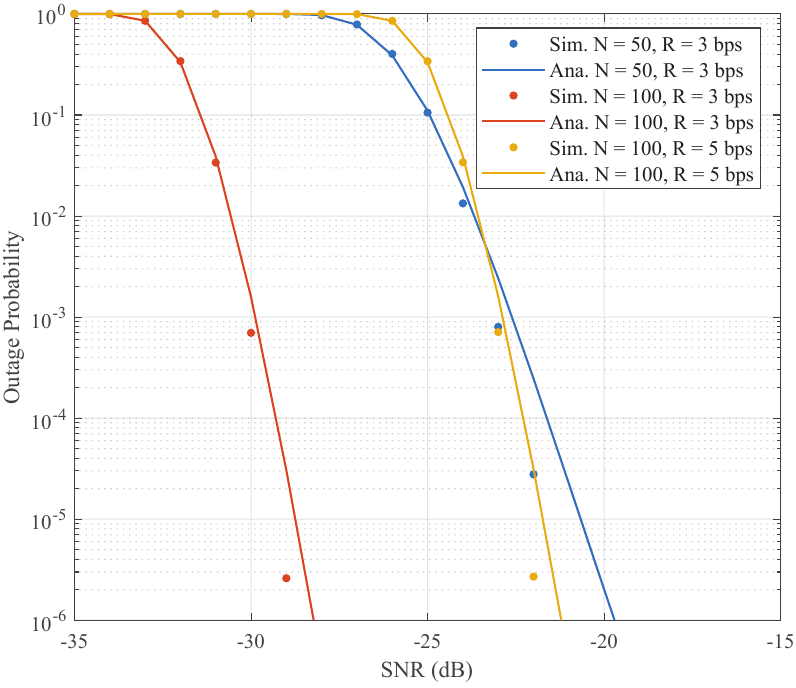}
\caption{\small{Outage probability analysis of the proposed systems.}}
\vspace{-10pt}
\label{OP1}
\end{figure}

In Fig. \ref{LI_16}, we plot the effect of LI on the proposed FD system under imperfect CSI, where we consider the estimation variance is a fixed value, i.e., $\sigma_e^2=0.1$.
Since the relatively small number of elements in RIS cannot satisfy the conditions of CLT, the analytical curve cannot be plotted. Thus, the results in Fig. \ref{LI_16} are obtained by Monte Carlo simulations.
To be specific, we can see from Fig. \ref{LI_16} that the performance of the proposed system becomes worse as the LI parameter $k_{AA}^2$ gets larger.
As the SNR increases, we observe that the ABEP curve will gradually tend to be stable in Fig. \ref{LI_16}(a). That is to say, when the SNR increases to a degree, the main factor affecting the performance is the channel estimation error rather than SNR, where the error floor can be effectively reduced by decreasing the value of $k_{AA}^2$.
In Fig. \ref{LI_16}(b), we observe that the overall ABEP curve is reduced compared to Fig. \ref{LI_16}(a). This indicates that the effect of LI on the RIS-FD-SSK system can be effectively counteracted by increasing the number of elements in the RIS.
From observing Figs. \ref{LI_16}(a) and \ref{LI_16}(b), we note that as the residual LI value $k_{AA}^2$ gets progressively smaller, the better the performance of the FD system.
Compared to the HD system, we find that the performance advantage of the FD system becomes obvious when the residual LI value becomes smaller in terms of ABEP. In addition, compared with Fig. \ref{LI_16}(a), we find that the performance of the FD system in Fig. \ref{LI_16}(b) is overall better than that of the HD system, which demonstrates that the increase in the number of RIS units can enhance the ABEP performance of the FD system.

\begin{figure*}[t]
 \centering
 \subfigure[{ Fixed $\sigma_e^2$}]
 {
  \begin{minipage}[b]{0.48\textwidth}
   \centering
   \includegraphics[width=8.7cm]{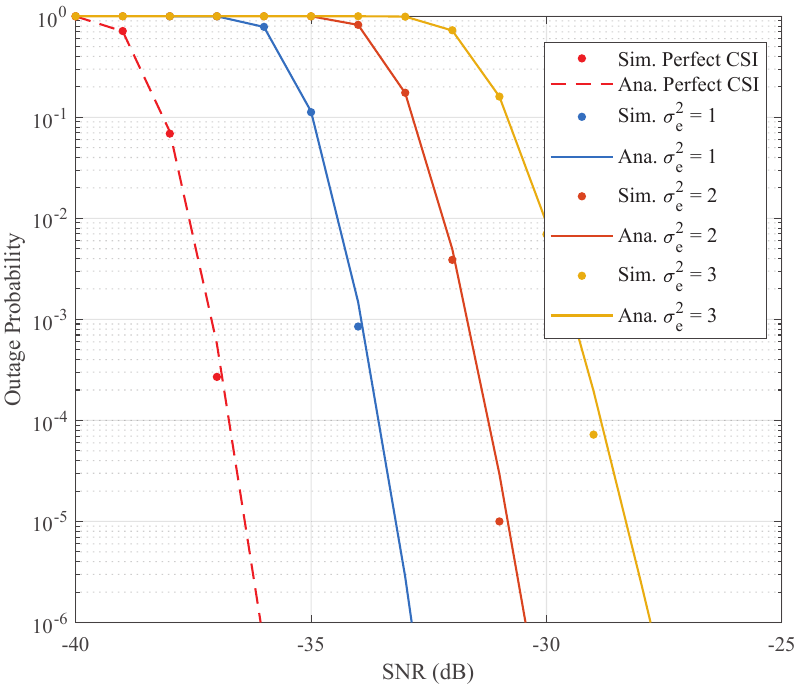}
  \end{minipage}
 }
 \subfigure[{ Variable $\sigma_e^2$}]
    {
     \begin{minipage}[b]{0.48\textwidth}
      \centering
      \includegraphics[width=8.7cm]{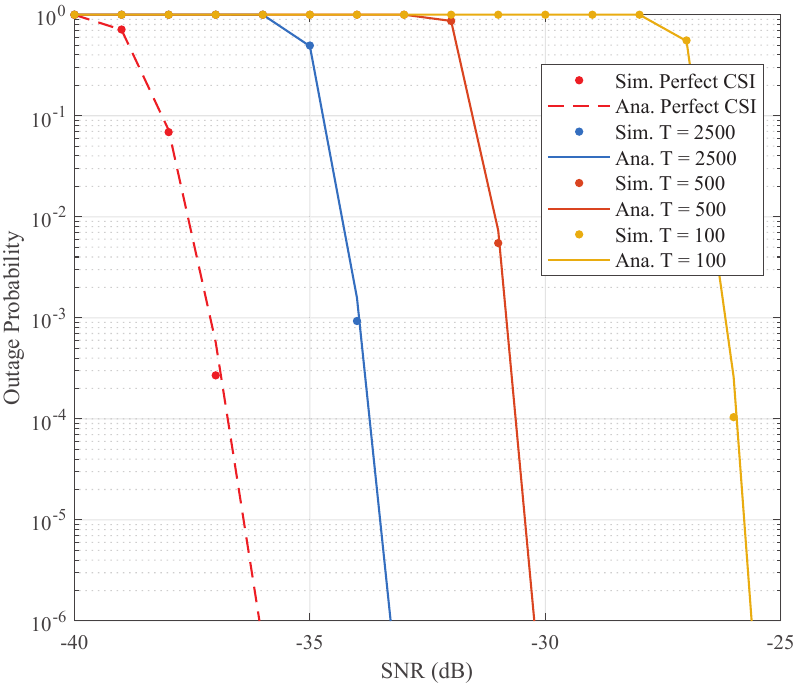}
     \end{minipage}
    }
\caption{\small{The impact of channel estimation error on the outage probability of the proposed systems.}}
\label{OP3}
\end{figure*}

\begin{figure*}[t]
 \centering
 \subfigure[Effect of $N$.]
 {
  \begin{minipage}[b]{0.48\textwidth}
   \centering
   \includegraphics[width=8.7cm]{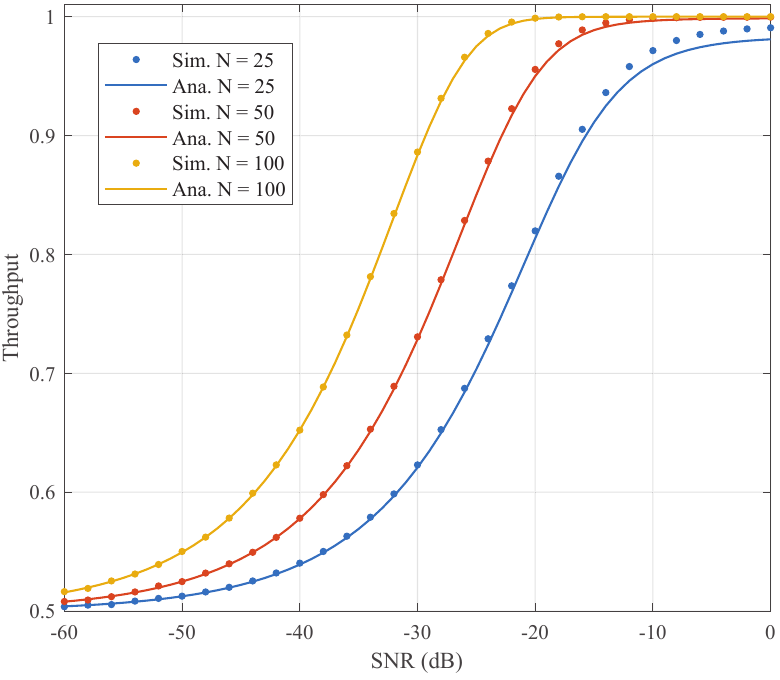}
  \end{minipage}
 }
 \subfigure[Effect of $N_t$.]
    {
     \begin{minipage}[b]{0.48\textwidth}
      \centering
      \includegraphics[width=8.7cm]{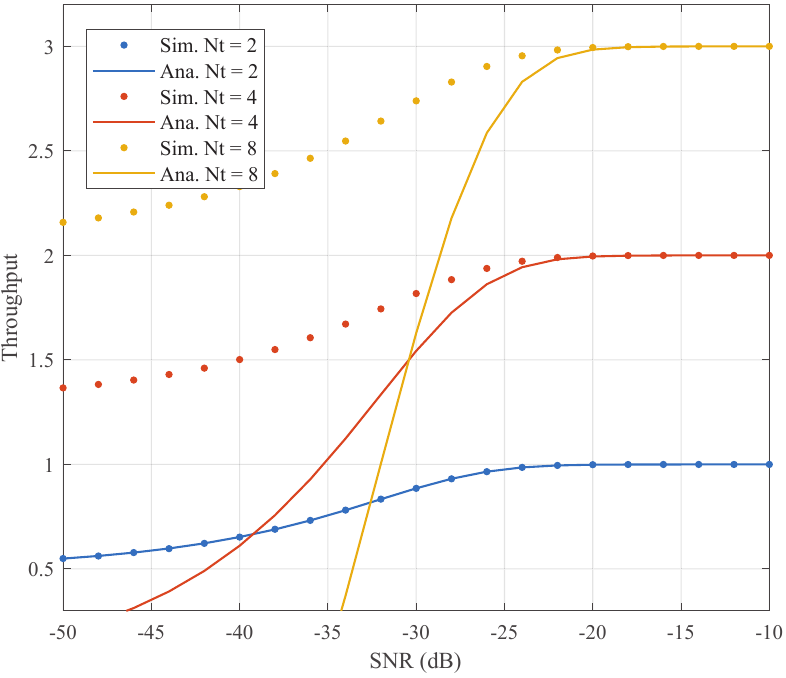}
     \end{minipage}
    }
\caption{\small{Throughput analysis of the RIS-FD-SSK systems.}}
\label{Th1}
\end{figure*}

Fig. \ref{contras_jin_high}(a) presents the ABEP of RIS-FD-SSK scheme with fixed channel estimation error as a function of the SNR, compared to benchmark schemes with HD scheme.
In particular, the LI parameter $k_{AA}^2$ is set as 0.3.
As expected, the simulation and analytical results for the FD system almost perfectly match the total number of reflection elements $N=400$. However, for the HD system, the simulation and analysis results match only in the high SNR region. This is because (\ref{abepx}) is the analytical result with $N_t=2$, while (\ref{abepx}) degenerates to the analytical tight upper bound expression when $N_t>2$, and this phenomenon also confirms the correctness of {\bf Remark 1}.
Furthermore, we observe that the performance difference between the FD and HD schemes is extremely small, while the FD scheme outperforms the HD scheme in the high SNR regions.

In Fig. \ref{contras_jin_high}(b), we compare the ABEP performance of RIS-FD-SSK with variable channel estimation error to that of the HD scheme over Rayleigh fading channels, where we set $k_{AA}^2$ as 0.3.
For the FD scheme, the analytical ABEP curves are precisely matched to the simulation curves in the whole SNR region.
It can be seen from Fig. \ref{contras_jin_high}(b) that the RIS-FD-SSK scheme shows a significant performance degradation compared to the case in the presence of channel estimation error.
We can observe from Fig. \ref{contras_jin_high}(b) that the FD scheme outperforms the HD scheme in the high SNR regions. In other words, the proposed scheme can provide better ABEP performance than the HD scheme even when the SNR is relatively larger.
In addition, when the number of pilots is larger, the performance with variable channel estimation error $\sigma_e^2$ is closer to the perfect CSI case.

In Fig. \ref{OP1},  we plot the outage probability results of the proposed scheme and study the impact of the number of reflecting elements and data rate on system performance.
It can be seen that as the number of reflecting elements is raised from 50 to 100, the outage probability curve is shifted to the left as a whole, which means that the performance of outage probability is improved.
This is because when the number of RIS elements is larger, the SINR becomes larger, which results in a lower outage probability value.
Additionally, the outage probability performance starts to deteriorate as $R$ changes from 3 bps to 5 bps, which is attributed to the increase in the threshold $\gamma_{th}$ as $R$ becomes larger.

In Fig. \ref{OP3}, we investigate the impact of channel estimation error on the outage probability performance of the RIS-FD-SSK scheme, where the number of elements $N$ and data rate $R$ of RIS are set to 200 and 3, respectively.
It is expected that the numerical simulation results match perfectly with the analytical results derived in (\ref{oplimi0}).
Note that Fig. \ref{OP3}(a) and Fig. \ref{OP3}(b) denote the cases where the channel estimation error $\sigma_e^2$ is a fixed parameter and a variable parameter, respectively.
We can observe that when the channel estimation error is smaller or more pilots are transmitted, the outage probability is closer to perfect CSI case.

Fig. \ref{Th1}(a) plots the throughput versus SNR for different numbers of RIS reflecting elements, where the channel estimation error variance $\sigma_e^2$ is set to a fixed value of one. It can be seen that when the number of RIS elements is higher, the simulation results and the analytical curves have a higher degree of agreement, which is due to the fact that the analytical expression is obtained based on the CLT. In addition, we find that as the SNR increases, the larger the number of RIS elements, the better throughput becomes accordingly. However, when the SNR increases to a certain level, throughput for different numbers of reflection elements tends to stable and close to one.

In Fig. \ref{Th1}(b), we show the effect of the number of transmit antennas on the throughput under the RIS-FD-SSK scheme with imperfect CSI, where the number of reflecting elements of the RIS and the error variance are set to 100 and 1, respectively.
According to Fig. \ref{Th1}(b), it is clear that in the high SNR region, the Monte Carlo simulation results tend to coincide with the analytical curves. However, when $N_t=2$, the two completely match in the considered SNR region. This is because the analytical upper bound is equivalent to the true value if $N_t=2$.
Furthermore, it is found that when $N_t$ is larger, it represents a higher modulation order. Simultaneously, the corresponding transmission rate is larger, and thus the throughput becomes greater.

\section{Conclusions}
This paper investigated the performance analysis concerning RIS-assisted FD two-way SSK communication system under imperfect CSI, where two FD users are equipped with corresponding RISs to enhance the signal transmission.
With the help of the ML detection algorithm and CLT, we derived the closed-form expression of ABEP obtained with channel estimation errors, which is obtained via the GCQ method. Then, we derived the upper bound and asymptotic closed-form expressions of the ABEP to gain more insights.
Furthermore, the outage probability and throughput of the considered system with imperfect CSI were also evaluated.
Additionally, we derived the outage probability and further provided the system throughput of the considered system.
We first evaluated the accuracy of the GCQ and the applicability conditions of the CLT.
Then, the analytical derivation results were validated using Monte Carlo simulations.
Lastly, we exhaustively studied the effect of all parameters on RIS-FD-SSK systems with imperfect CSI.
Results demonstrate that the FD system outperforms the HD system when the residual LI is small or the number of units in the RIS is large in terms of ABEP. In addition, the FD system can achieve better ABEP performance than the HD system in the high SNR region for both fixed and variable channel estimation errors.
In the future, one can focus on the impact of hybrid RIS on the proposed system, and the fundamental trade-off between the spectral efficiency and reliability of the system can be explored, which are promising and open research directions.

\end{document}